\documentclass[a4paper,10pt]{article}
\pdfoutput=1 
\usepackage{jheppub} 
\usepackage{lineno}
\usepackage{orcidlink}
\usepackage{hhline}
\usepackage{slashed}
\usepackage{graphicx}
\usepackage{graphics}
\usepackage{dcolumn}
\usepackage{bm}
\usepackage{amssymb}
\usepackage{enumerate}
\usepackage{lineno}
\usepackage{multirow} 
\usepackage{color}
\usepackage{hyperref}
\usepackage{xcolor}
\usepackage{booktabs}
\usepackage{soul}

\usepackage{footnote}
\makesavenoteenv{tabular}
\makesavenoteenv{table}

\newcommand*{\Apr}{{A^\prime}}
\newcommand*{\pot}{$e^+$OT }

\title{Proof of principle for a light dark matter search with low-energy positron beams at NA64}
\collaboration{NA64 collaboration}
\collaborationImg{\includegraphics[width=0.2\textwidth]{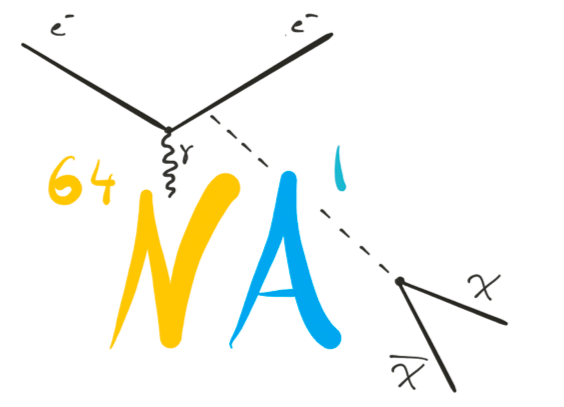}}

\author[a]{Yu.~M.~Andreev\orcidlink{0000-0002-7397-9665}}
\author[b]{A.~Antonov\orcidlink{0000-0003-1238-5158}}
\author[c,d]{M.~A.~Ayala~Torres\orcidlink{0000-0002-4296-9464}}
\author[e]{D.~Banerjee\orcidlink{0000-0003-0531-1679}}
\author[f]{B.~Banto Oberhauser\orcidlink{0009-0006-4795-1008}}
\author[g]{V.~Bautin\orcidlink{0000-0002-5283-6059}}
\author[e]{J.~Bernhard\orcidlink{0000-0001-9256-971X}}
\author[b,h]{P.~Bisio\orcidlink{/0009-0006-8677-7495}}
\author[i]{M.~Bondì\orcidlink{0000-0001-8297-9184}}
\author[b]{A.~Celentano\orcidlink{0000-0002-7104-2983}}
\author[e]{N.~Charitonidis\orcidlink{0000-0001-9506-1022}}
\author[f]{P.~Crivelli\orcidlink{0000-0001-5430-9394}}
\author[a]{A.~V.~Dermenev\orcidlink{0000-0001-5619-376X}}
\author[a]{S.~V.~Donskov\orcidlink{0000-0002-3988-7687}}
\author[a]{R.~R.~Dusaev\orcidlink{0000-0002-6147-8038}}
\author[g]{T.~Enik\orcidlink{0000-0002-2761-9730}}
\author[g]{V.~N.~Frolov}
\author[g]{S.~V.~Gertsenberger\orcidlink{0009-0006-1640-9443}}
\author[e]{S.~Girod}
\author[a]{S.~N.~Gninenko\orcidlink{0000-0001-6495-7619}}
\author[l]{M.~H\"osgen}
\author[g]{Y.~Kambar\orcidlink{0009-0000-9185-2353}}
\author[a]{A.~E.~Karneyeu\orcidlink{0000-0001-9983-1004}}
\author[g]{G.~Kekelidze\orcidlink{0000-0002-5393-9199}}
\author[l]{B.~Ketzer\orcidlink{0000-0002-3493-3891}}
\author[a]{D.~V.~Kirpichnikov\orcidlink{0000-0002-7177-077X}}
\author[a]{M.~M.~Kirsanov\orcidlink{0000-0002-8879-6538}}
\author[a,g]{V.~A.~Kramarenko\orcidlink{0000-0002-8625-5586}}
\author[a]{L.~V.~Kravchuk\orcidlink{0000-0001-8631-4200}}
\author[a,g]{N.~V.~Krasnikov\orcidlink{0000-0002-8717-6492}}
\author[c,d]{S.~V.~Kuleshov\orcidlink{0000-0002-3065-326X}}
\author[d]{V.~E.~Lyubovitskij\orcidlink{0000-0001-7467-572X}}
\author[g]{V.~Lysan\orcidlink{0009-0004-1795-1651}}
\author[b,1]{A.~Marini\note{Corresponding author.}\orcidlink{0000-0002-6778-2161}}
\author[b]{L.~Marsicano\orcidlink{0000-0002-8931-7498}}
\author[g]{V.~A.~Matveev\orcidlink{0000-0002-2745-5908}}
\author[d]{R.~Mena~Fredes}
\author[d,m]{R.~Mena~Yanssen}
\author[n]{L.~Molina Bueno\orcidlink{0000-0001-9720-9764}}
\author[f]{M.~Mongillo\orcidlink{0009-0000-7331-4076}}
\author[g]{D.~V.~Peshekhonov\orcidlink{0009-0008-9018-5884}}
\author[a]{V.~A.~Polyakov\orcidlink{0000-0001-5989-0990}}
\author[o]{B.~Radics\orcidlink{0000-0002-8978-1725}}
\author[g]{K.~Salamatin\orcidlink{0000-0001-6287-8685}}
\author[a]{V.~D.~Samoylenko}
\author[f]{H.~Sieber\orcidlink{0000-0003-1476-4258}}
\author[a]{D.~Shchukin\orcidlink{0009-0007-5508-3615}}
\author[d,p]{O.~Soto}
\author[a]{V.~O.~Tikhomirov\orcidlink{0000-0002-9634-0581}}
\author[a]{I.~Tlisova\orcidlink{0000-0003-1552-2015}}
\author[a]{A.~N.~Toropin\orcidlink{0000-0002-2106-4041}}
\author[n]{M.~Tuzi\orcidlink{0009-0000-6276-1401}}
\author[a,g]{P.~V.~Volkov\orcidlink{0000-0002-7668-3691}}
\author[a]{I.~V.~Voronchikhin\orcidlink{0000-0003-3037-636X}}
\author[c,d]{J.~Zamora-Sa\'a\orcidlink{0000-0002-5030-7516}}
\author[g]{A.~S.~Zhevlakov\orcidlink{0000-0002-7775-5917}}

\affiliation[a]{Authors affiliated with an institute covered by a cooperation agreement with CERN}
\affiliation[b]{INFN, Sezione di Genova, 16147 Genova, Italia}
\affiliation[c]{Center for Theoretical and Experimental Particle Physics, Facultad de Ciencias Exactas, Universidad Andres Bello, Fernandez Concha 700, Santiago, Chile}
\affiliation[d]{Millennium Institute for Subatomic Physics at High-Energy Frontier (SAPHIR), Fernandez Concha 700, Santiago, Chile}
\affiliation[e]{CERN, European Organization for Nuclear Research, CH-1211 Geneva, Switzerland}
\affiliation[f]{ETH Z\"urich, Institute for Particle Physics and Astrophysics, CH-8093 Z\"urich, Switzerland}
\affiliation[g]{Authors affiliated with an international laboratory covered by a cooperation agreement with CERN}
\affiliation[h]{Universit\`a degli Studi di Genova, 16126 Genova, Italia}
\affiliation[i]{INFN, Sezione di Catania, 95123 Catania, Italia}
\affiliation[l]{Universit\"at Bonn, Helmholtz-Institut f\"ur Strahlen-und Kernphysik, 53115 Bonn, Germany}
\affiliation[m]{Universidad T\'ecnica Federico Santa Mar\'ia and CCTVal, 2390123 Valpara\'iso, Chile}
\affiliation[n]{Instituto de Fisica Corpuscular (CSIC/UV), Carrer del Catedratic Jose Beltran Martinez, 2, 46980 Paterna, Valencia, Spain}
\affiliation[o]{Department of Physics and Astronomy, York University, Toronto, ON, Canada}
\affiliation[p]{Departamento de Fisica, Facultad de Ciencias, Universidad de La Serena, Avenida Cisternas 1200, La Serena, Chile}

\emailAdd{anna.marini@ge.infn.it}

\abstract{
Thermal light dark matter (LDM) with particle masses in the 1 MeV - 1 GeV range could successfully explain the observed dark matter abundance as a relic from the primordial Universe. In this picture, a new feeble interaction acts as a ``portal'' between the Standard Model and LDM particles, allowing for the exploration of this paradigm at accelerator experiments. In the last years, the ``missing energy'' experiment NA64$e$  at CERN SPS (Super Proton Synchrotron) has set world-leading constraints in the vector-mediated LDM parameter space, by exploiting a 100 GeV electron beam impinging on an electromagnetic calorimeter, acting as an active target.
In this paper, we report a detailed description of the analysis  of a preliminary measurement with a 70 GeV/c positron beam at NA64$e$, performed during summer 2023 with an accumulated statistics of $1.596 \times 10^{10}$ positrons on target (hereafter referred to as \pot). This data set was analyzed with the primary aim of evaluating the performance of the NA64$e$ detector with a lower energy positron beam, towards the realization of the post-LS3 program. The analysis results, other than additionally probing unexplored regions in the LDM parameter space, provide valuable information towards the future NA64$e$ positron campaign.
}

\begin{document} 
\maketitle
\flushbottom
\section{Introduction} 

Despite being the most comprehensive theory of Nature at our disposal, the Standard Model (SM) of particle physics still does not include Dark Matter (DM), one of the  unsolved questions of modern physics~\cite{Bertone:2016nfn,Liddle:1998ew,Arcadi:2017kky,Arbey:2021gdg}. Among the several DM models proposed over the years, thermal light dark matter (LDM), assuming DM particles $\chi$ with masses below the electroweak scale, has recently aroused significant interest in the particle physics community. In this class of models, the currently observed DM density in the Universe arises as the relic of a primordial era, when DM was in thermal equilibrium with the SM particles. According to this hypothesis, a new interaction is necessary to cope with the current DM abundance. A possibility assumes that the interaction between the $\chi$ particles and the SM is realized by a new massive vector boson, usually called ``dark photon'' or ``$\Apr$''~\cite{Holdom:1985ag,Izaguirre:2013uxa,Batell:2009di}. The $\Apr$, with a sub-GeV mass, may interact with the SM via kinetic mixing with the ordinary photon, and equally act as the mediator of a new ``dark'' force, being associated to a spontaneously broken $\rm U(1)_D$ dark gauge group. The low-energy effective Lagrangian describing the $\Apr$ and LDM reads (omitting the $\chi$ mass term):
\begin{equation}
    \mathcal{L} \supset -\frac{1}{4} F'_{\mu\nu} F'^{\mu\nu} +\frac{m_{A'}^2}{2} A'_\mu A'^\mu +\frac{\varepsilon}{2} F'_{\mu\nu} F^{\mu\nu} + g_D A'_\mu J_\chi^\mu.
\end{equation}

Here $F'_{\mu\nu}=\partial_\mu A'_\nu -\partial_\nu A'_\mu$ and $ F_{\mu\nu}$ are, respectively, the dark photon and SM electromagnetic field strength, $m_{A'}$ is the $A'$ mass, $\varepsilon$ determines the strength of the $A-A'$ kinetic mixing, $g_D = \sqrt{4 \pi \alpha_D}$ is the dark gauge coupling and $J_\chi^\mu$ is the current of LDM particles. In this picture, after diagonalization of the kinetic part of the underlying Lagrangian the mixing term $\frac{\varepsilon}{2} F_{\mu\nu}' F^{\mu\nu}$ generates the interaction Lagrangian ${\cal L}_{int} = \varepsilon e A_\mu' J^\mu_{em}$ between the dark photon and electromagnetic current $J^\mu_{em}$ composed of SM fields, i.e. SM charged particles acquire an effective coupling $\varepsilon e$ with the $\Apr$.  Origin of the kinetic mixing between visible and dark photons can be explained by loop processes involving particles charged under both SM $U_{em}(1)$ and $U_{D}(1)$ Abelian gauge groups~\cite{Essig:2010ye,delAguila:1988jz,Arkani-Hamed:2008kxc}.  

Taking into account these loop effects leads to an estimate of $\varepsilon \sim 10^{-4} - 10^{-2}$ (at one-loop order) and $\varepsilon \sim 10^{-6} - 10^{-3}$ (at two-loops order).

In the thermal hypothesis of DM, the currently observed DM density can be correlated with the parameters of the model. Thus, it is possible to define values of the parameter $y$, proportional to the velocity-mediated DM annihilation cross section:
\begin{equation}
    y = \alpha_D \varepsilon^2  \left(\frac{m_\chi}{m_\Apr} \right)^4 \propto \langle \sigma_{ann.} v \rangle m_{\chi}^2
\end{equation}
by imposing the freeze out of the relic DM density, providing a target for discovery or rejection of the  model~\cite{Battaglieri:2017aum}. More precisely, the following constraint on $\alpha_D$ can be derived~\cite{Berlin:2018bsc}:
\begin{equation}
    \alpha_D \simeq 0.02 \cdot f \left( \frac{10^{-3}}{\varepsilon}\right)^2 \left(\frac{m_\Apr}{100 \,\, \rm MeV}\right)^2 \left(\frac{10 \, \rm MeV}{m_\chi}\right)^4, 
\end{equation}
where $f$ is a dimensionless parameter depending on the theoretical details of the models: in the representative case $\frac{m_\chi}{m_\Apr}=\frac{1}{3}$, $f \lesssim 10 \, (1)$ for a scalar (fermion) LDM candidate $\chi$.
This scenario gives rise to a rich particle phenomenology, requiring the complementarity between different experimental techniques towards a final discovery or disproval of this hypothesis. Depending on the $\frac{m_\chi}{m_\Apr}$ mass ratio, the dark photon features two main decay modes: if  $m_\Apr > 2\, m_\chi$, then, due to the small LDM-SM coupling, the $\Apr$ predominantly decays to a $\chi \bar{\chi}$ pair, in the so-called \textit{invisible} decay scenario. Otherwise, if there is no ``dark'' state lighter than the $\Apr$, it decays \textit{visibly} to SM states. Fixed target experiments at the intensity frontier are particularly well suited to explore this scenario, not suffering from any kinematic suppression for the detection of light DM and being less susceptible to the form factor induced cross
section suppression which affects direct LDM searches for  specific models~\cite{Essig:2011nj}. A review of current and proposed measurements can be found in ~\cite{Battaglieri:2017aum,Essig:2013lka,Alexander:2016aln,Beacham:2019nyx,Antel:2023hkf}. In this work we will restrict ourselves to the invisible decay scenario, presenting the result of a preliminary LDM search with a positron beam in the fixed-target experiment NA64$e$ at the CERN Super Proton Synchrotron (SPS)~\cite{NA64:2019imj,Andreev:2021fzd}.

\section{The NA64$e$ experiment}

\subsection{The missing energy technique}
Among the different experimental strategies adopted to search for an $\Apr$ invisibly decaying to LDM particles, the \textit{missing-energy} technique 
has proven to be particularly efficient~\cite{Fabbrichesi:2020wbt,Filippi:2020kii,Beacham:2019nyx,Battaglieri:2017aum,Ilten:2022lfq,Graham:2021ggy}. In a missing-energy experiment, a high energy electron/positron beam impinges on a thick, active target, where the developing electromagnetic shower is fully contained, and the energy deposited is measured. In the event of LDM production from the interaction of the particles in the shower with the target, the ``dark'' particles would escape from the detector, carrying away a significant fraction of the initial beam energy. In this scenario, LDM production is clearly signaled by a large missing energy, i.e. the difference between the energy of the incoming particle of the beam and that one deposited in the active target. Since this experimental technique relies only on the $\chi$ production and does not require the LDM particles scattering in the detector, it features enhanced sensitivity, for a given collected statistic, with respect to traditional beam-dump experiments. 
Indeed, for a missing-energy experiment the signal yield scales as $\varepsilon^2$, while for beam-dumps the different signal signature results in a $\varepsilon^4$ $\alpha_D$ scaling.
In an $e^+/e^-$ beam experiment, two main LDM production processes contribute to the total signal yield~\cite{Marsicano:2018glj,Marsicano:2018krp}: the so-called $\Apr$-strahlung, a radiative process analogous to the standard bremsstrahlung ($e^\pm Z \rightarrow e^\pm Z \Apr$), followed by the $\Apr$ decay to a $\chi\bar{\chi}$ pair, and the resonant annihilation of an energetic positron with an electron in the target ($e^+e^-\rightarrow \Apr \rightarrow \chi\bar{\chi}$). Resonant annihilation may occur both with a positron or an electron beam; in the latter case,  only 
the secondary positrons of the electromagnetic shower  produced in the active target contribute to this production channel. Due to the $s$-channel kinematics of the process, for a narrow $\Apr$ ($\alpha_D \lesssim 0.1$), the resonant annihilation cross-section is rapidly suppressed as the invariant mass of the interacting $e^+e^-$ pair deviates from $m_\Apr^2$,  resulting in the $\Apr$-strahlung being the dominating LDM production process in a large $m_\Apr$ range, at a given beam energy. However, if kinematically accessible, the resonant annihilation dominates the LDM production yield, due to its favorable scaling $\alpha_{EM} Z$ w.r.t. the $\alpha_{EM}^3 Z^2$ scaling of $\Apr$-strahlung~\cite{Marsicano:2018glj} (here Z is the atomic number of the target material). In addition, resonant annihilation is not affected by  the loss of nuclear coherence and by
the reduction of the Weizäcker-Williams effective photon flux limiting the $\Apr$-strahlung yield at large $A'$ masses. 
In missing-energy experiments, the number of signal events from LDM production via resonant annihilation scales approximately as:

\begin{equation}
N_{signal|res. ann} \propto \int_{E_{th}}^{E_{beam}} \frac{dT_+}{dE}(E) \sigma_{res.}(E) dE,
\label{eq:N_res}
\end{equation}
where $E_{beam}$ is the beam energy, $E_{th}$ is the missing-energy threshold defining the signal, $\frac{dT_+}{dE}$ is the differential track length\footnote{The positrons track-length distribution $dT_+(E)/dE$ is
defined as the integral over the active-target volume of the differential fluence $\Phi(E)$, corresponding to the density of particle tracks in the volume \cite{Chilton}. Intuitively, the quantity $dT_+(E)$ represents the path length in the target taken by positrons with energy in the interval between $E$ and $E+dE$.} of positrons in the target with respect to the positron energy $E$, and $\sigma_{res.}$ is the resonant annihilation cross section. 

\begin{figure}[t]
    \centering
    \includegraphics[width=0.5\linewidth]{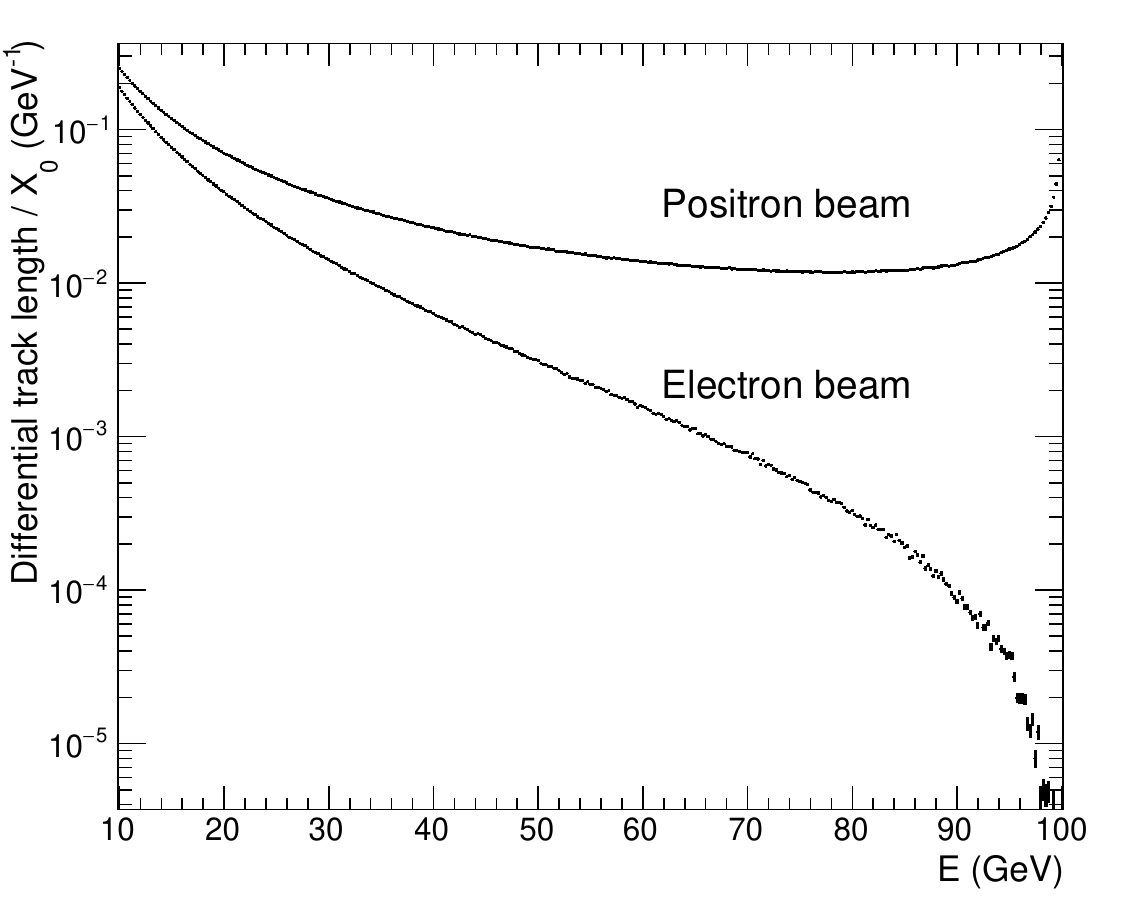}
    \caption{Differential track length $\frac{dT_+}{dE}$ of positrons in a thick target, for a 100 GeV electron and positron beam. In the electron case, only secondary $e^+$ contribute to $\frac{dT_+}{dE}$, resulting in a monotone, decreasing track length, dominated by low energy positrons; on the other hand, an $e^+$ beam results in a significantly larger $\frac{dT_+}{dE}$, peaked at $E_{beam}$, due to the primary particle contribution.}
    \label{fig:track-length}
\end{figure}

As shown in Fig.~\ref{fig:track-length}, the shape and magnitude of the differential track length differs radically whether the experiment exploits an $e^-$ or an $e^+$ beam. 
It follows from Eq.~\ref{eq:N_res} that, for a narrow $\Apr$, one has $N_{signal|res. ann}\propto T_+(E_{res}) \cdot \sigma_{res.}(E_{res})$ where $E_{res}\simeq \frac{m_\Apr^2}{2 m_e}$ is the resonant energy in the laboratory frame, provided that $E_{th} \lesssim E_{res}\lesssim E_{beam}$. As a consequence, a missing-energy experiment can fully exploit the resonant annihilation contribution only to search for $\Apr$s with masses in the range: 
\begin{equation}\label{eq:A_mass_range}
\sqrt{2 m_e E_{th}} \lesssim m_{\Apr} \lesssim \sqrt{2 m_e E_{beam}}.
\end{equation}
These considerations motivates the realization of a multi-energy, positron-beam, missing energy experiment, allowing to explore the largest portion of the LDM parameter space exploiting the resonant annihilation process. The NA64$e$ experiment at CERN SPS represent the ideal experimental environment where to realize such an experimental program.

\subsection{Searching for LDM with positrons at NA64$e$} \label{sec:NA64}

\begin{figure}[t]
    \centering
    \includegraphics[width=0.95\linewidth]{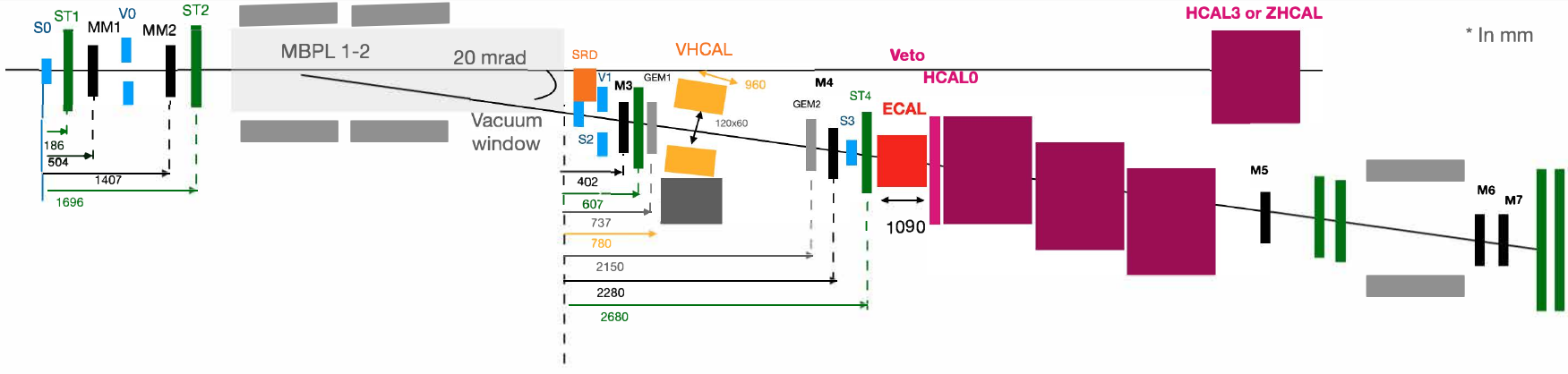}
    \caption{Overview of the NA64 experimental setup for the 2023 data-taking.}
    \label{fig:NA642023setup}
\end{figure}

The missing-energy technique for the search of invisibly decaying particles was pioneered by the NA64$e$ experiment at the CERN SPS (Super Proton Synchrotron). The experiment, exploiting the $\mathcal{O}(100 \,\, \rm GeV)$ $e^+/e^-$ high-purity beam at the H4 line of CERN North Area, has been operating since 2016.  A schematic view of the NA64$e$ experimental setup, for the 2023 configuration, is shown in Fig.~\ref{fig:NA642023setup}. 
At H4, the 70 GeV/c positron beam is obtained by having the primary 400 GeV/c proton beam from SPS impinging on a thick Be target, and pair-converting the emerging forward photon with a thin Pb foil. A downstream septum magnet is used to select the momentum and charge of the particles transported toward the detector, located approximately 500 m downstream.
The North Area facility at CERN receives the 400 GeV/c proton beam through a slow extraction process, resulting in a “spill-like” structure. Each spill lasts for 4.8 seconds and is followed by a 10/15-second interval~\cite{Roncarolo:2852569}. The typical beam intensity on the Be target per SPS spill is about $10^{10}$ protons-on-target.

In NA64e, the incoming beam is tagged by a set of three plastic scintillator counters (S0, S1, S2) and two veto counters (V0, V1). A magnetic spectrometer, made by a set of tracking detectors (GEMs, MicroMegas, and Straw tubes) installed upstream and downstream of a two dipole magnets with total magnetic strength $\int B dl \simeq 7\,$T$\,\times\,$m~\cite{Banerjee:2017mdu}, is used to measure their momentum, with typical resolution $\delta p/p$ of about $1\%$~\cite{Banerjee:2017mdu}.
The MM1 and MM2 detectors define the upstream part of the track relative to the magnet spectrometer, while MM3, MM4, GEM1, and GEM2 measure the downstream segment.
To suppress the residual hadronic contamination, a compact Pb/Sc calorimeter (SRD) is installed just after the magnet to detect the synchrotron radiation emitted by positrons along the curved trajectory in the magnetic field region~\cite{Depero:2017mrr}. 
The NA64 active target is a 40$X_0$ Pb/Sc Shashlik calorimeter (ECAL), assembled as a $5\times6$ matrix of $3.82\times3.82$ cm$^2$ cells, each coupled with a PMT for an independent readout. The calorimeter is longitudinally divided into a $4X_0$ pre-shower section and a main section. In front of the ECAL, a prototype of a veto-hadronic calorimeter (VHCAL), made by Cu/Sc layers with a 12$\times$ 6 cm$^2$ hole in the middle, is installed, to suppress the dominant background from large-angle hadronic secondaries produced by the interaction of the primary beam with upstream materials. A full-scale VHCAL detector, with optimized angular coverage, is currently matter of an R\&D program within the NA64 collaboration~\cite{vhcalBenjamin}. Downstream the ECAL, a hermetic Fe/Sc hadronic calorimeter (HCAL) is installed. It consists of three modules with a total length of approximately 30$\lambda_I$. The HCAL is designed to detect secondary hadrons and muons produced in the ECAL. A high-efficiency plastic scintillator counter (Veto) is installed between the ECAL and the HCAL to further reduce backgrounds. Last, a fourth HCAL module is installed at zero degrees to detect neutral particles generated by the beam's interaction with the upstream beam line materials.

The main trigger for the experiment, adopted in production runs, is provided by the coincidence of the $S0$ and $S2$ counters, in anti-coincidence with $V0$ and $V1$, and further requiring for an in-time energy deposit in the ECAL below 52 GeV to align with the ``missing energy'' signal signature -- this value was optimized as a trade-off between the requirement of a small missing-energy threshold to optimize the experiment sensitivity and the maximum trigger rate sustainable by the DAQ system.
Additionally, the energy deposited in the first ECAL module (pre-shower) must exceed a threshold of 200 MeV. A pre-scaled ``calibration trigger'' is also implemented, solely based on the $S$ and $V$ counters, for calibration and monitoring. In this manuscript, the events acquired solely with the ``calibration trigger'' are referred to as ``calibration'' events or runs, in contrast to ``production" events or runs, where the ECAL trigger thresholds are applied.

The NA64 positron program started during summer 2022, with a test run collecting $\sim 10^{10}$ 100-GeV \pot. This data proved the potential of the resonant annihilation process at NA64$e$, strongly motivating a dedicated program at lower beam energies~\cite{NA64:2023ehh}. This program, consisting of two main measurements with 40-GeV and 60-GeV positron beams, was proposed to CERN Super Proton Synchrotron Experiments Committee (SPSC)~\cite{Bisio:2887649}, with the aim to start data taking after CERN \textit{Long Shutdown 3}.
The operation of the NA64$e$ detector with lower beam energies presents however experimental challenges to be addressed, one above all, the hermeticity of the detector at $E_{beam}<100$ GeV needs to be studied in detail.
In this perspective, during summer 2023, the collaboration performed a second test run with a positron beam, lowering the particle momentum to 70 GeV/c. In the next sections, we present the physics results of the analysis of 2023 70 GeV/c positron data taking, focusing on the technical details and highlighting the implications towards the future multi-energy program  proposed to SPSC. 

\section{The 70 GeV/c positron measurement}

The presented analysis is based on a total statistics of $1.596\times 10^{10}$ \pot,
accumulated during 24 production runs, with a nominal intensity of about $4.2\times10^6$ particles per
SPS spill of 4.8 s.
In this configuration, at 70 GeV/c nominal beam momentum, the expected residual hadronic contamination in the H4 beam is $\eta_h =(0.457 \pm 0.007)\%$, mainly from  protons produced by the $\Lambda \rightarrow p \pi^-$ decay~\cite{Andreev:2023xmj}. 
This section reports the fundamental elements of the analysis, starting from the description of the data processing at the detector level and moving to the upper level analysis; the results are discussed in Sec.~\ref{sec:res70GeV}.

\subsection{Accumulated charge measurement}
The luminosity, i.e. the total number of impinging particles on the detector, is measured by counting with a scaler system the number of signals satisfying the $S0-S2-\overline{V}0-\overline{V}1$ coincidence matrix, in anti-coincidence with the DAQ system BUSY signal.
We observe that the same coincidence matrix enters in the trigger condition for the production runs. Furthermore, Monte Carlo (MC) simulations for signal events are normalized to this measured value. Hence, any uncertainty in the system, for example due to a fluctuation in the response of any of the scintillator counters, factorizes out in the determination of the final upper limit. 

\subsection{Data processing}\label{sec:dataProc}

\begin{figure}[t]
    \centering
    \includegraphics[width=.32\textwidth]{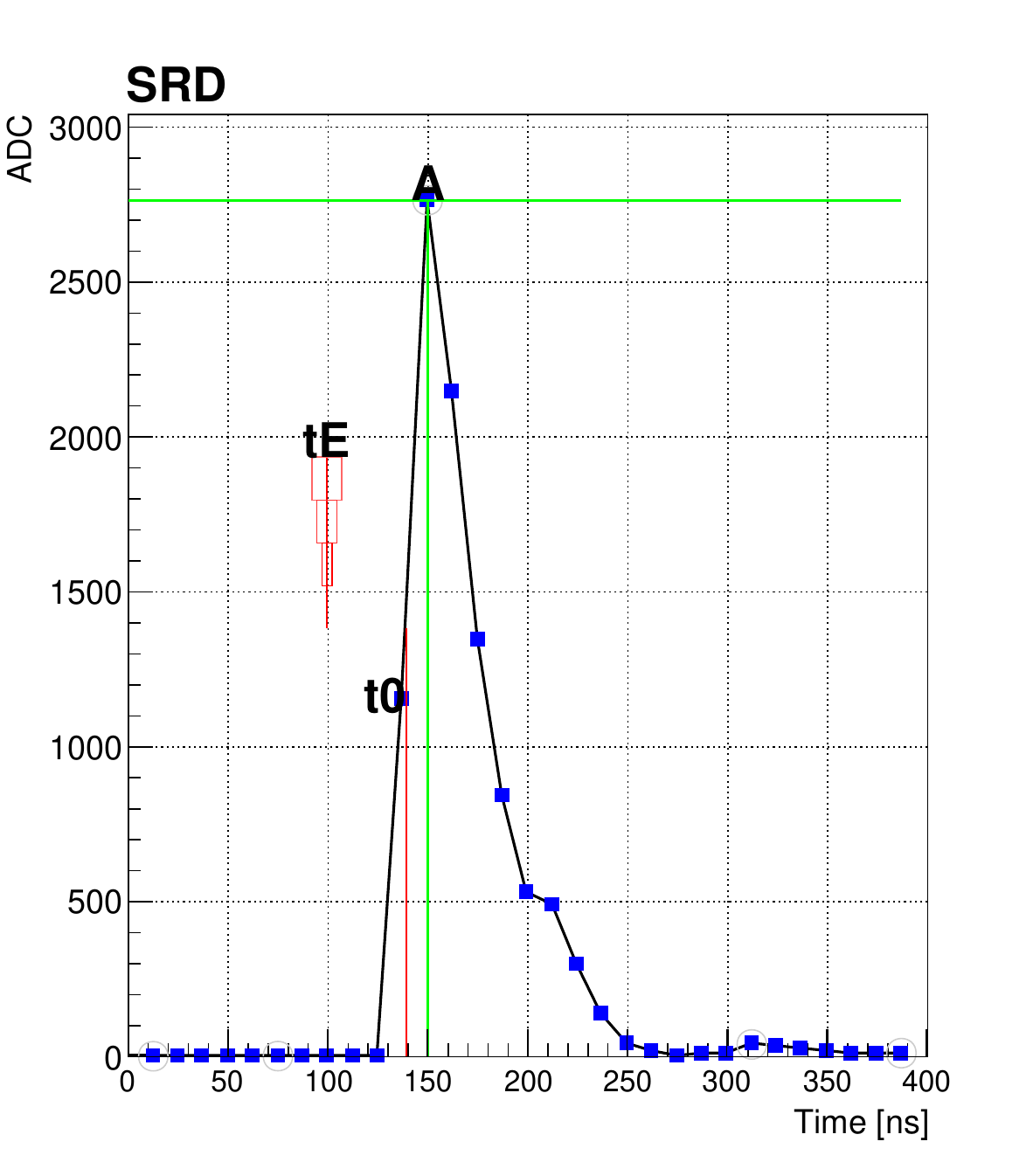}
    \includegraphics[width=.32\textwidth]{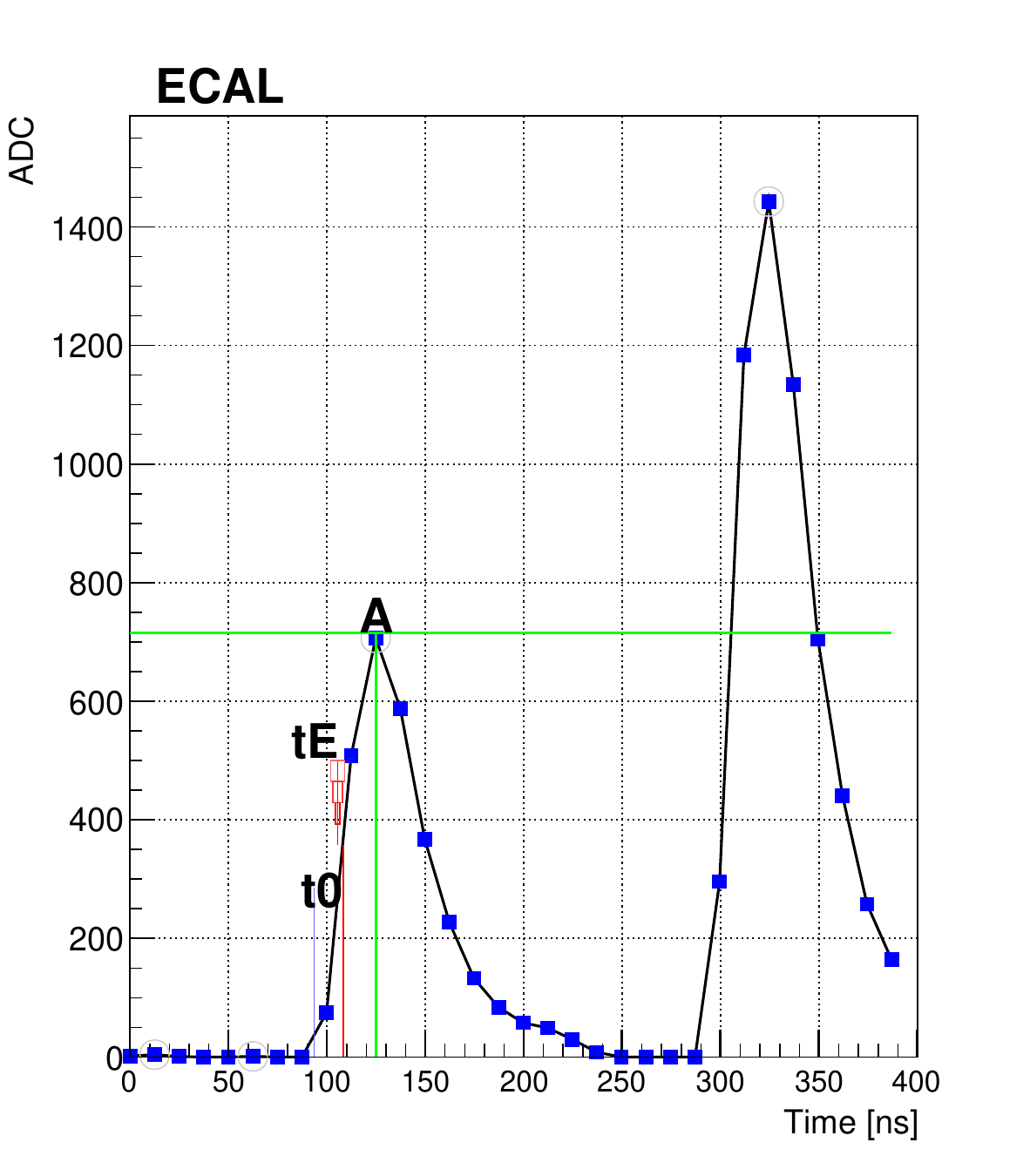}
    \includegraphics[width=.32\textwidth]{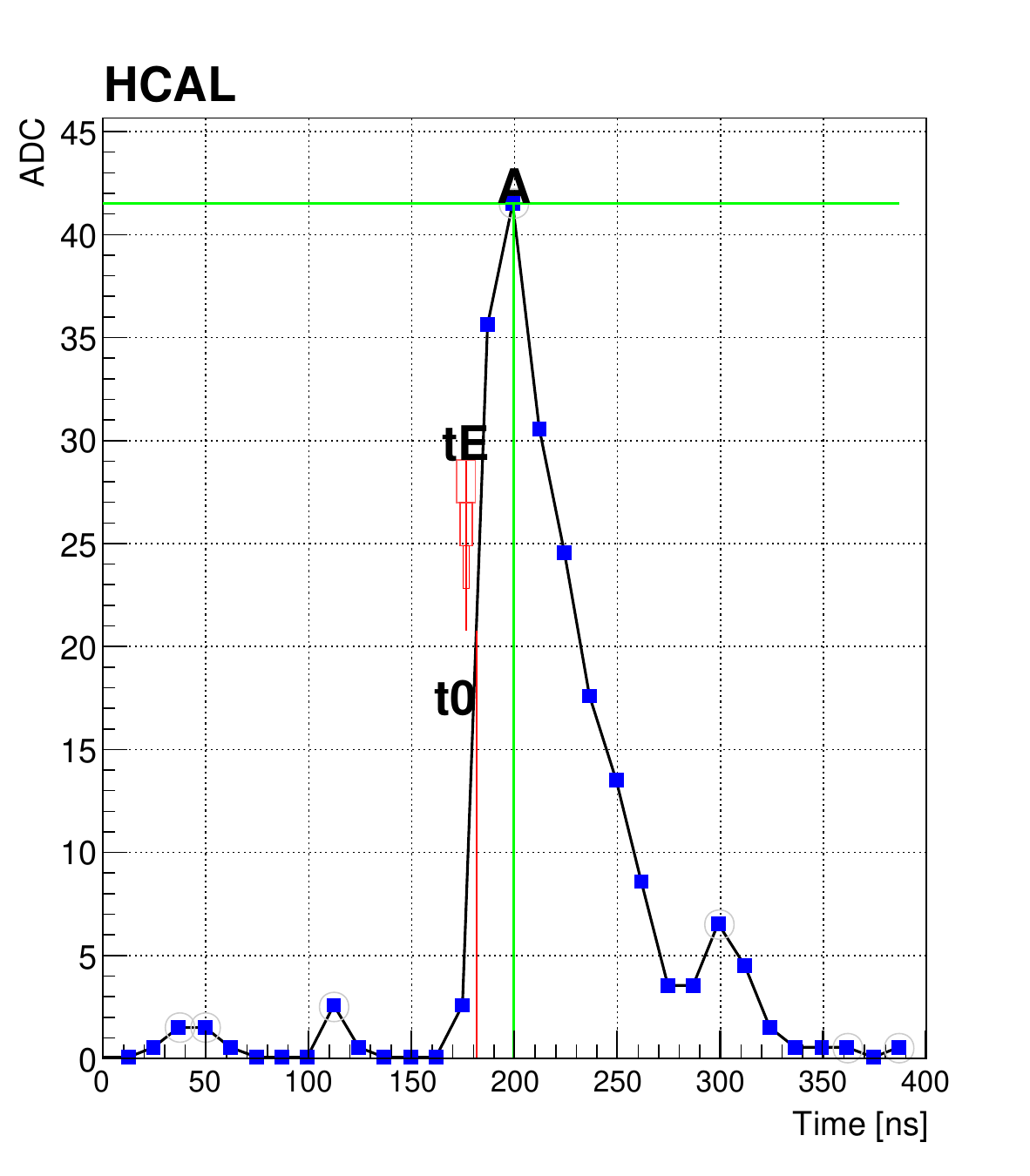}
    \caption{Illustration of a typical pile-up event measured during the NA64-2023 $e^+$ run. The left, middle, right plot show, respectively, the measured waveform for the central SRD cell, the central ECAL cell, and the central HCAL0 cell. In each plot, the vertical red line corresponds to the measured peak time $t_0$, while the three red rectangles are centered at the expected time $t_E$ and correspond, respectively, to the $\pm1$ $\sigma_{t_E}$, $\pm2$ $\sigma_{t_E}$, $\pm3$ $\sigma_{t_E}$ intervals. The ECAL signal clearly shows a double-peak structure, with a first small in-time pulse, followed by a larger one. The HCAL signal (right) shows a single in-time pulse. Finally, the SRD signal (left) also shows a single pulse, that is, however, at a much larger time than the expected one. This event can be explained as the superposition of a proton impinging first on the detector, followed immediately after by a positron. The proton passed through the ECAL with a small energy deposit therein and then released all its energy in the HCAL, with no associated activity in the SRD. The positron, instead, resulted in a visible energy deposit in the SRD, and a large signal in the ECAL.}
    \label{fig:reco1}
\end{figure}

For all PMT-based detectors (SRD, ECAL, VETO, HCAL), the analogue signal of each cell is digitized via a 12 bit, 80 MHz ADC, implementing a waveform shaping filter at the front-end. For each recorded event, the 32 acquired samples in a 400-ns long time window aligned with the trigger time were saved. First, for each event and each cell, the waveform pedestal is evaluated from the average of the first ten samples. After pedestal subtraction, individual peaks in the waveform are resolved (a peak is defined for each sample in the waveform higher than two neighborhood ones). For each peak the hit energy $E_i$ is reconstructed by taking the amplitude of the sample and applying an appropriate calibration constant, while the hit time $t_i$ is obtained from the pulse leading edge through a constant-fraction like extrapolation algorithm.
Due to the large beam intensity ($\simeq~1$ MHz) 
a small fraction of the events is associated with two beam particles impinging on the detector almost at the same time, with the corresponding PMT signals superimposing within the same readout window\footnote{Considering the typical NA64 detector response time of $\sim 50$ ns, the probability for two particles to be partially superimposed is about 5\% for a beam intensity of 1 MHz.}. This results in multiple peaks associated to the same waveform. To mitigate this effect, a pile-up suppression algorithm was implemented. For each cell, the expected signal time $t_E$ and its uncertainty $\sigma_{t_E}$ were first evaluated from calibration runs and stored in the calibration database. Then, during reconstruction, for each waveform the peak closer to the expected time $t_E$ were selected. The superposition of different particles in the same event, characterized by different interaction mechanisms with the detector, may also induce spurious, out-of-time effects in some detector cells. For example, in case of a proton impinging on the detector just before a positron, the SRD waveform is characterized by a single peak, located at a late time with respect to the proton-induced trigger (see Fig.~\ref{fig:reco1} for an example). To compensate for this, the following time-cut definition was applied to each measured energy:
\begin{equation}
    E_{in-time} = 
    \left\{
    \begin{array}{cl}
    E & \mathrm{if}\,\, |t_0-t_E|<n\cdot \sigma_{t_E} \\
    0 & \mathrm{otherwise}
    \end{array}
    \right.
    \; \; ,
    \label{eq:intime}
\end{equation}
where $t_0$ is the time of the peak closer to $t_E$. In this analysis, the value $n=3$ was used for SRD, VHCAL, and ECAL, while for VETO and HCAL $n=5$ was chosen to maximize the rejection power. We observe that, other than mitigating the pile-up effect, the in-time algorithm reduces the effect of the electrical noise for cells with zero energy deposit. The corresponding reconstructed energy is set to zero any time the noise-induced peak closer to the expected time does not satisfy the condition $|t_0-t_E|<n\times\sigma_{t_E}$.

For each channel of the tracking detectors (Micromegas and GEMs), the APV25-based readout system provides three analog charge samples at intervals of 25 ns~\cite{French:2001xb}. This data is then used for cluster reconstruction and hit-point determination. First, the pedestal for each electronic channel is measured by collecting data without beam impinging on the setup. These pedestals are defined as the mean of the charge samples for each channel, with the corresponding RMS value ($\sigma$) indicating the noise level per channel. When the beam impinges on the detector, a strip is considered active if its third charge sample exceeds the pedestal by at least 3$\sigma$. The particle hit position is reconstructed using a mapping of channel strips, accounting for multiplexing in the case of Micromegas, and a clustering algorithm that groups signals from adjacent channels. Further details of the reconstruction algorithm can be found in~\cite{Banerjee:2017mdu}.

\subsubsection{Calorimeters calibration}

\begin{figure}
    \centering
    \includegraphics[width=0.45\linewidth, height=0.45\linewidth]{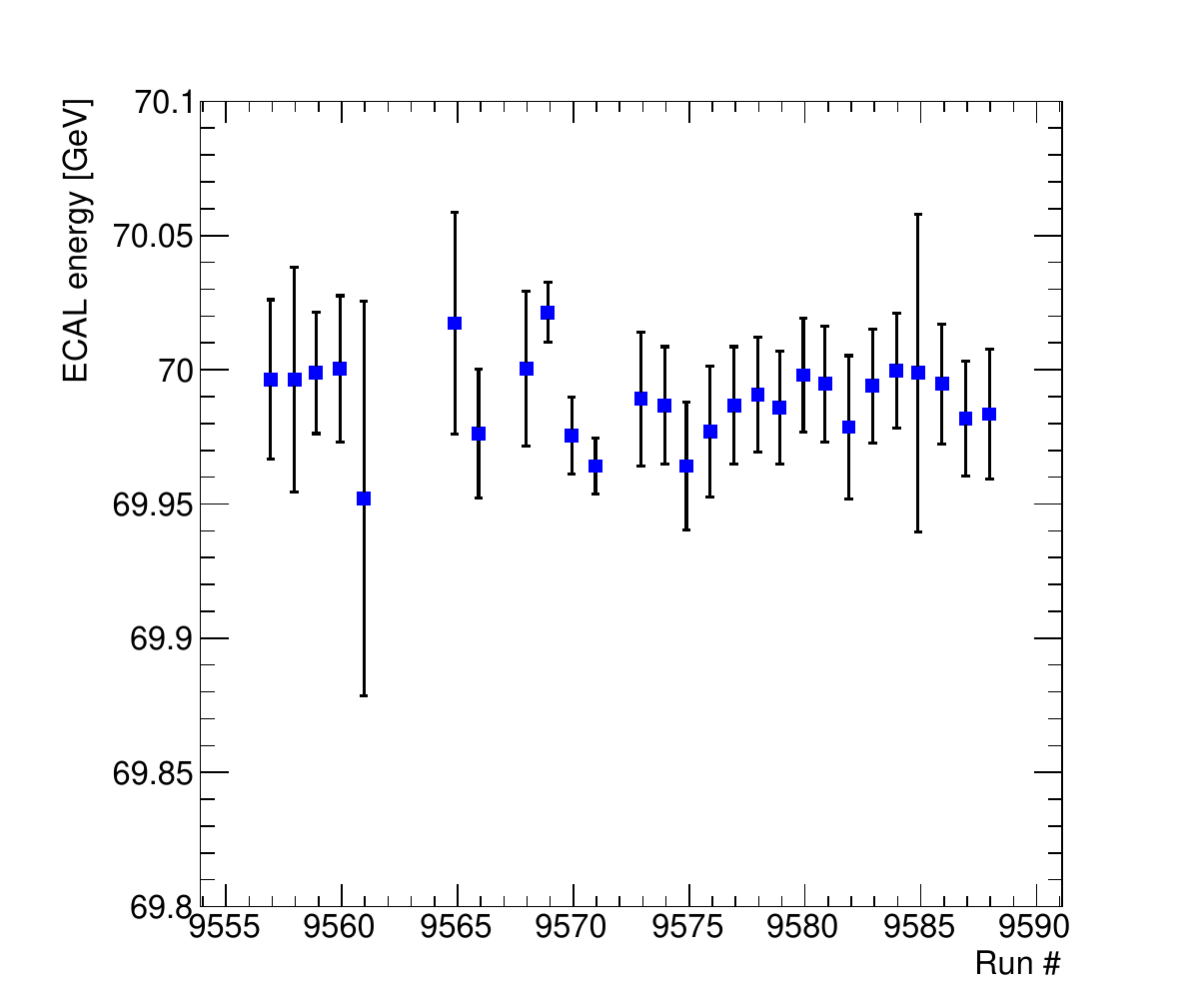}
    \includegraphics[width=0.45\linewidth]{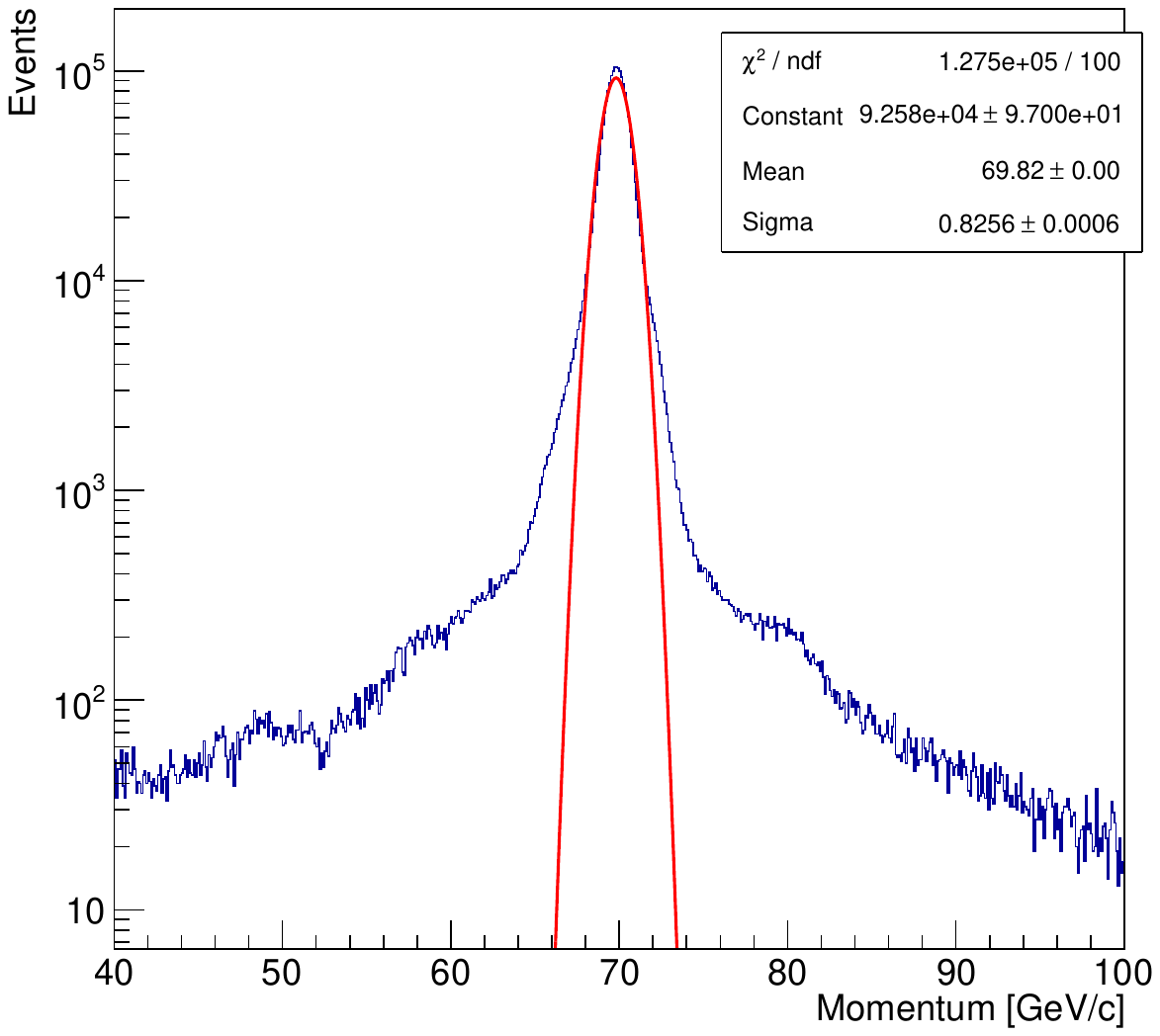}
    \caption{Left panel: average energy deposit in the ECAL for events recorded by the prescaled ``calibration-trigger'' condition, for each production run. For each data point, the error bar corresponds to a $\pm 1\sigma$ for the corresponding distribution. Right panel: the reconstructed momentum distribution from tracking detectors for all the calibration events. The distribution was fitted with a Gaussian function.}
    \label{fig:calib}
\end{figure}

The energy response of each ECAL cell was calibrated through the data collected during a set of calibration runs, in which the position of the detector was changed to have the beam impinging in the center of a given $(x,y)$ cell. For each of these, we started from a first-order value of the calibration constants determined by considering the run in which the beam impinged on it, and comparing the average signal amplitude with the predicted energy deposit in the cell from simulations. Then, we adopted an iterative procedure, considering again for each cell the corresponding run and selecting events with a clean impinging $e^+$ on the ECAL, by requiring the presence of a well-reconstructed track and large-enough energy deposit in the SRD detector. For these events, we measured the energy deposit in the pre-shower section and in the main section, extracting the corresponding average values $E^{x,y}_{ECAL-0}$ and $E^{x,y}_{ECAL-1}$. These were compared with the predictions from MC, to obtain a correction to the cell calibration constants:
\begin{align}
c^{x,y}_{ECAL-0} &= \left(E^{x,y}_{ECAL-0}\right)_{MC} /\left(E^{x,y}_{ECAL-0}\right)_{data} \\
c^{x,y}_{ECAL-1} &= \left(E^{x,y}_{ECAL-1}\right)_{MC} /\left(E^{x,y}_{ECAL-1}\right)_{data}\;\;.
\end{align}
Since the values of $E^{x,y}_{ECAL-0}$ and $E^{x,y}_{ECAL-1}$ depend on the calibration factors of all cells, the procedure was iterated until $c^{x,y}\simeq 1$. 
Finally, to correct for long-term variations of the detector response, we introduced specific spill-dependent correction factors, determined by measuring the relative variation of each cell signal with respect to that from the first run. For the central cell, the correction factor was based on the analysis of the events collected during production runs with the prescaled ``calibration'' trigger, while for periphery cells reference signals induced from calibration LEDs were used. 
After the calibration, the stability of the ECAL was assessed by comparing the average ECAL energy deposit for events recorded by the prescaled ``calibration-trigger'' condition, for each production run, and found to be within $\simeq 0.1\%$ (see also Fig.~\ref{fig:calib}, left panel).

To calibrate the response of the HCAL, VETO, and VHCAL detectors, instead, we adopted a simpler procedure, based on data collected after configuring the H4 beam line to transport a muon beam with a large transverse size. We measured the deposited energy in each cell, featuring a Minimum Ionizing Particle (MIP) distribution, and compared it with the prediction from a MC simulation to extract the calibration constant. This approach is justified by the fact that, in the present analysis, these detectors are basically used as a ``veto'' systems, and no spectral information is extracted from them. 

\subsubsection{Momentum reconstruction}\label{sec:momReco}

The reconstruction of the impinging-particle track relies on hit information provided by MM and GEM detectors. 
A fitting procedure based on the GenFit framework~\cite{Rauch:2014wta} provides track parameters such as the reconstructed momentum, angles before and after the magnet, fitting $\chi^2$ and $p$-value (see also Fig.~\ref{fig:calib} right-panel). GenFit supports various track-fitting techniques, among which the Deterministic Annealing Filter (DAF) algorithm showed the best robustness and track reconstruction efficiency, making it the preferred choice for data processing.

\subsection{Analysis strategy and selection criteria} \label{sec:cuts}

The data analysis consisted in the choice and optimization of the selection criteria defining a signal event, through a multi-step procedure based both on data and MC simulations.  A ``blind analysis'' approach was adopted, by defining a signal window in the 2D space of the  ECAL and HCAL in-time energy deposit.  The blinded signal region was delimited by the following condition, requiring a missing-energy of at least 28 GeV and negligible activity in the HCAL:
\begin{equation}
    E_{ECAL}<42~\mathrm{GeV}, \, E_{HCAL} < 1~\mathrm{GeV} \; \; .
\end{equation}
The selection criteria based on the information collected by the different NA64$e$ sub-detectors were optimized with the general prescription of maximizing the signal efficiency and simultaneously tuning the background rejection power of the cuts. Since, as showed in Sec.~\ref{sec:backgrounds}, the backgrounds evaluation for such an experiment is subject to large systematic uncertainty -- due to the inherent nature of the measurement -- the  requirement of keeping the value of the expected background in the signal window $B_{exp} \ll 1$ was adopted. Given the non-trivial correlations between the significant number of variables, corresponding to the different sub-detectors, collected for each event, we decided to optimize the signal selection with a systematic approach, dealing sequentially with each detector. As described in detail in the following sections, different strategies for the  optimization were followed, depending on the nature of the selection: where 
possible, the cut definition was guided by physical considerations, e.g. the VETO energy cut was defined below the energy value corresponding to the passage of a minimum ionizing particle (MIP). In other cases, the optimization was performed by evaluating the cut impact on both signal and expected backgrounds.
 Analysis cuts can be categorized in three groups: ``upstream'', those relative to all the sub-detector upstream the active target, ``ECAL'', and ``downstream'' cuts, involving the VETO and HCAL. In the subsequent sections, we will delve into the significance and role of each of the applied selections. 

\subsubsection{Upstream cuts}\label{sec:upstreamCuts}
\begin{figure}[t]
    \centering
    \includegraphics[width=0.515\linewidth]{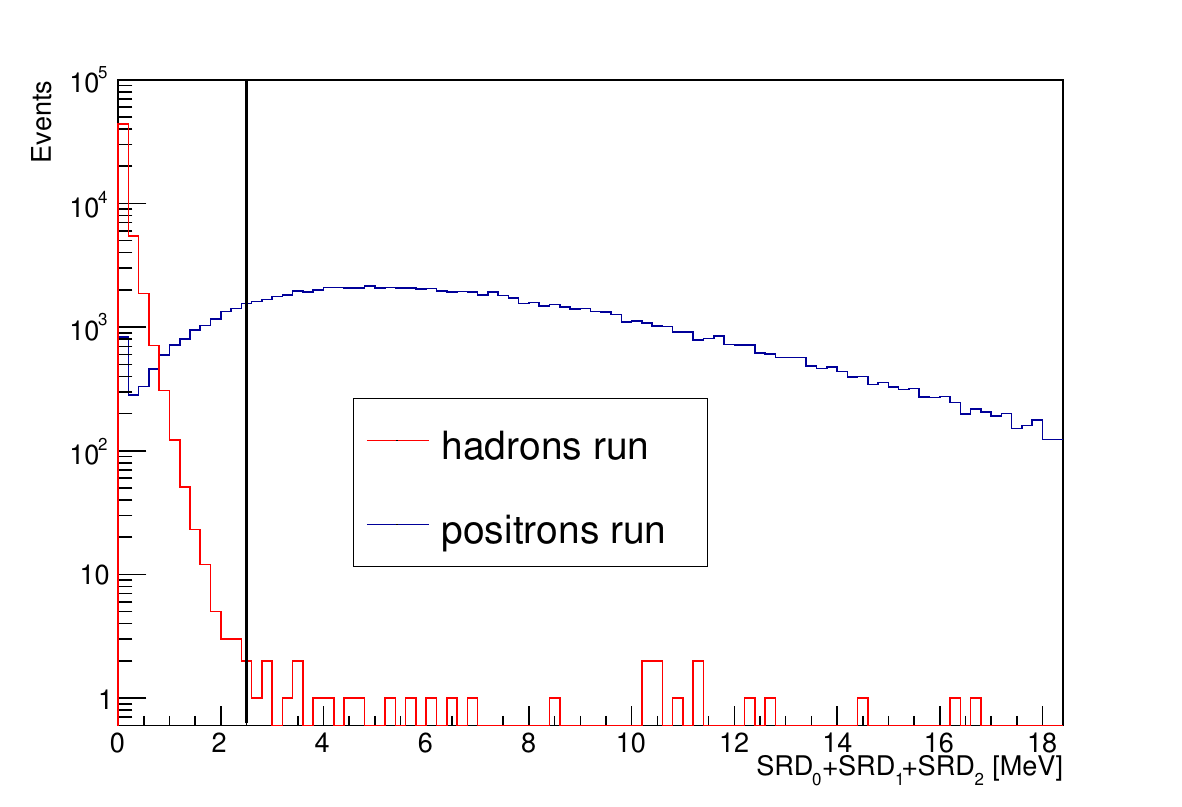}
    \includegraphics[width=0.465
\linewidth]{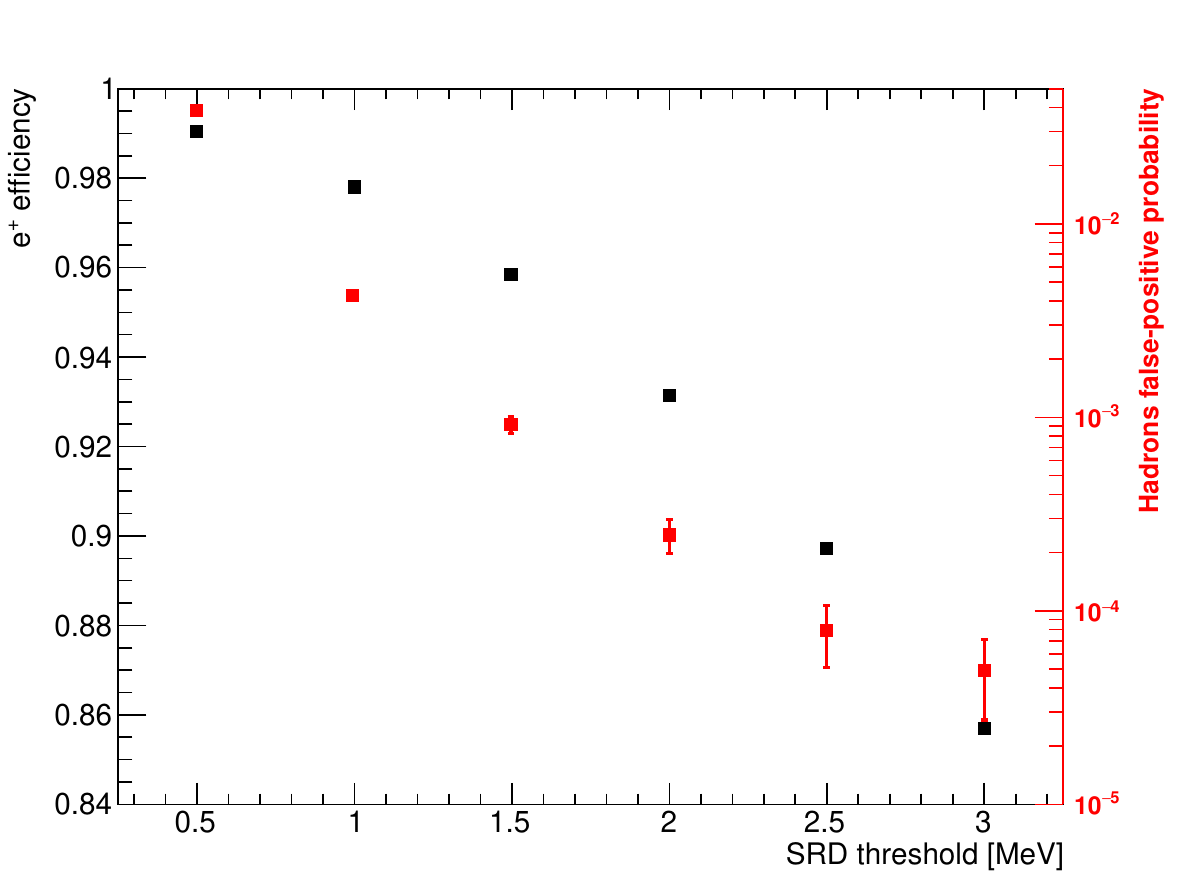}
    \caption{Left: total energy deposited in all SRD modules by positron (blue) / hadron (red) events. The vertical black line corresponds to the 2.5 MeV cut employed in the analysis. Right: SRD detector efficiency for impinging 70 GeV/c $e^+$ (black) / false-positive probability for hadrons (red) as a function of the energy threshold -- for most of the data points, the size of the error bar is smaller than the marker.}
    \label{fig:SRDspectrum}
\end{figure}

The aim of the upstream selection is to identify incident positrons with an energy of 70 GeV and to reject all other particles, including lower-energy positrons and hadronic contaminants of the beam~\cite{Andreev:2023xmj}. The detectors involved in this selection are mainly the SRD, the tracking system, the VHCAL and the Straw Tubes. 
For the optimization of the SRD,  tracking, and  VHCAL cuts we exploited the calibration events, identifying a clean set of 70 GeV
positron events through a procedure not involving the cut under study. We then applied it to the
selected sample to investigate the corresponding relative efficiency as a function of the threshold
value. Specifically, we identified impinging positrons requiring that the total energy deposited in the ECAL was close to 70 GeV, with less than 10 GeV deposited in the HCAL,
and no activity in the VETO detector. 
For the momentum selection, we defined the acceptance window as [P$_0$ - 2$\sigma$, P$_0$ + 2$\sigma$], where P$_0$ and $\sigma$ are the average beam momentum and the corresponding standard deviation obtained from a Gaussian fit of the momentum distribution (see Fig.~\ref{fig:calib}, right panel). The same approach was adopted to define the track angle cut, which is required to be less then 3 mrad.
We set a 2 MeV threshold on the SRD total energy, resulting in a 93\% detection efficiency for positron events and a false-positive rate of 2.5$\times10^{-4}$, obtained from the study of dedicated hadron calibration runs. Fig.~\ref{fig:SRDspectrum}  reports the SRD energy spectra for positron and hadron runs (left panel) and the efficiency/false-positive probability of the selection (right panel). For further details on the background rejection power of the SRD see Sec.~\ref{sec:InFlightDecay}.
Given the sensitivity of the VHCAL response
to the beam trajectory, the event sample used for the corresponding efficiency study was chosen
with the additional requirement of a well reconstructed impinging track, with momentum in the range 65 - 75 GeV/c. Ideally, no significant activity, apart from low-energy particle splash-back and electrical noise, is expected in the VHCAL for a clean positron track; the analysis of the calibration events resulted in a 1.5 GeV threshold on the total energy deposited across all VHCAL cells (see Sec.~\ref{sec:upstreamInteractions} for details on the role of this detector in background rejection).
Finally, the Straw Tubes cut definition was based on previous studies, revealing that the multiplicity of hits associated with events giving rise to upstream interactions is typically greater than five. Consequently, we required the multiplicity of hits in each Straw detector to be equal or less than this value.


\subsubsection{ECAL cuts}\label{sec:downstream_cuts}
These cuts include the ``topological'' ECAL cuts, based on the distribution of the energy deposit in the different ECAL cells, and pile-up cuts. For a signal consistent with $\Apr$ production, most in-time energy must be deposited in the central ECAL cell (EC(2,2) cut), and the following condition must be satisfied:
\begin{equation}
    \frac{E_{5\times6}^{ECAL} - E_{3\times3}^{ECAL} }{ E_{5\times6}^{ECAL}}  < 0.06 \; \;,
\end{equation}
where $E_{5\times6}^{ECAL} $ is the total ECAL energy, while $E_{3\times3}^{ECAL}$ is the energy deposited in the $3\times3$ cells central matrix. 
As mentioned in Sec.~\ref{sec:NA64}, the NA64$e$ ``production'' trigger features a pre-shower threshold of 200 MeV. To prevent potential efficiency fluctuations due to ECAL calibration imperfections, a more conservative pre-shower cut of 350 MeV, safely above the trigger threshold, was applied in the analysis. 
A further, more sophisticated algorithm, the ``shower-shape'' cut,  is applied to reject events with a different ECAL topology from the typical electromagnetic shower, indicating the possibility of an hadronic interaction in the ECAL. The selection is based on the so called $\chi^2_{shower}$ variable, defined as follows:
\begin{equation}
    \chi^2_{shower} = \sum_{i=0}^{N_{cells}}\left(\frac{f^{exp}_i(x_{meas},y_{meas}) - f^{meas}_i}{e_i(x_{meas},y_{meas})}\right)^2,
\end{equation} 
where $f^{exp}_i(x,y)$ is the expected energy fraction deposited in the $i-$th cell, for a given beam impact point $(x,y)$ on the ECAL, $f^{meas}_i$ is the measured energy fraction for the considered event and $e_i$ is the error associated to the prediction of $f^{exp}_i(x,y)$. The value of $f^{exp}_i(x_{meas},y_{meas})$, as well as the error $e_i$, are obtained from a reference electromagnetic shower shape ``profile'', built from 70 GeV/c calibration data.
The numerical value of the $\chi^2_{shower}$ cut was defined by studying how the quantity is distributed for ``di-muon'' events, where a muon pair is produced in the ECAL via the processes $e^- Z \rightarrow \mu \bar{\mu} e^- Z$ or $e^+ e^- \rightarrow \mu \bar{\mu} $ . 
This class of events is, from the point of view of the ECAL energy deposit, approximately analogous to signal events, since high-energy muons deposit negligible energy, which results in a significant missing-energy as in the LDM production case\footnote{For a complete discussion on the difference between the signal and ``di-muon'' event topology, see Appendix~\ref{app:chi2_correction}}. Therefore, di-muons provide a class of signal-like events that can be used as a benchmark to optimize the $\chi^2_{shower}$ cut, limiting the bias introduced by a MC based procedure.
\begin{figure}[t]
    \centering
    \includegraphics[width=.7\textwidth]{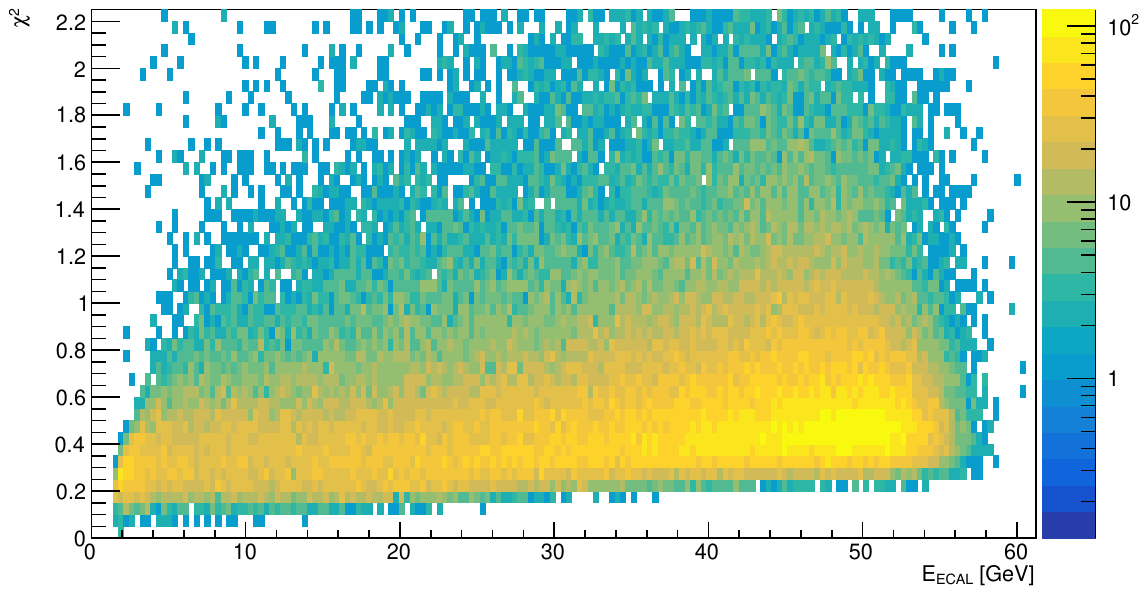}
    \caption{2D distribution of the  $\chi^2$ versus the measured energy in the ECAL, for the collected di-muon events.}
    \label{fig:dimuonchi2}
\end{figure}
Fig.~\ref{fig:dimuonchi2} shows the 2D distribution of the computed  $\chi^2_{shower}$ versus the ECAL energy deposit, for a sample of $\sim10^5$ di-muon events collected during the whole data taking (see Sec.~\ref{sec:ECALSignalMC} for a description of the di-muon events selection procedure in production-data). In order to optimize the signal efficiency of the cut, independently from the ECAL energy deposit, a set of different cut values $\chi^2_{{\rm cut}-i}$, obtained as follows, was defined. First, the di-muon sample was divided into 11 sub-samples, corresponding to 5-GeV-wide ECAL energy bins, from 0 to 55 GeV; for each $i-$th subset, the $\chi^2_{{\rm cut}-i}$ value was set so that applying the condition  $\chi^2_{shower} < \chi^2_{{\rm cut}-i}$ resulted in a fixed efficiency $\eta_{\chi^2}=95\%$ for di-muons events in that subset. The obtained $\chi^2_{{\rm cut}-i}$ values ranged in the interval (0.6-2.1). 

The last ECAL cut is aimed at rejecting events affected by pile-up, possibly interfering with the correct energy deposit measurement. This cut asks for a difference between the out-of-time and in-time ECAL energy below 30 GeV. 

\subsubsection{Downstream cuts}
\begin{figure}
    \centering
    \includegraphics[width=0.7\linewidth]{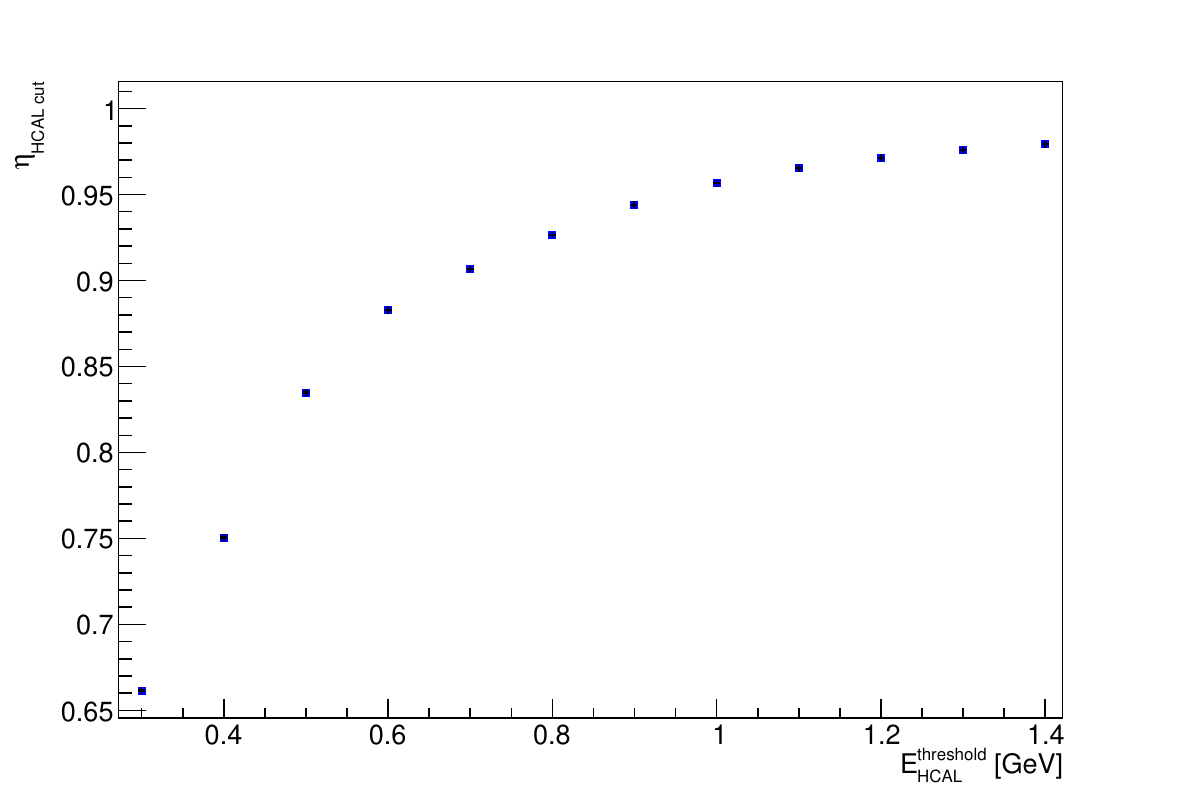}
    \caption{Efficiency of the HCAL cut, defined as $E_{HCAL} < E_{HCAL}^{threshold}$, where $E_{HCAL}$ is the total energy deposited in the three modules, as a function of the threshold applied to the total energy deposited in the first three modules. The error bars on the $y$ axis are covered by the marker size.}
    \label{fig:HCALeff}
\end{figure}
The downstream cuts, applied to the VETO and hadron calorimeters, play the role of rejecting events with a significant energy leaking from the ECAL, due to the production of SM penetrating particles, both charged and neutral. The HCAL cut requires the total energy deposited in the first three modules to be lower than a threshold value  $E_{HCAL}^{threshold}$, while the VETO cut is satisfied if the measured energy in each panel is simultaneously lower than a threshold value $E_{VETO}^{threshold}$. As for the upstream case, we relied on a clean sample of impinging 70 GeV/c positron events from calibration runs to optimize the cuts, using them as a template of the signal. This procedure is slightly conservative since, for a signal event, where an energetic LDM particle pair is produced, the probability of a significant energy deposit in the VETO and HCAL due to leakage is lower, since less energy is available for the production of SM penetrating particles. Still, we preferred to adopt a data-driven approach rather than a MC-based procedure, in order not to introduce in the optimization any bias due to eventual MC inaccuracies. Fig.~\ref{fig:HCALeff} shows the HCAL cut efficiency for the selected sample as a function of the energy threshold applied: a 1-GeV threshold was selected, resulting in a $\sim 95\%$ efficiency. An analogous procedure resulted in a 6 MeV threshold per Veto module (we underline that a MIP deposits on average $\sim10$ MeV in a VETO Module).

\subsection{Combined signal efficiency} \label{sec:sig_eff}

The detection efficiency of the whole analysis cut chain for LDM signals $\eta_\Apr(m_\Apr)$ can be expressed as the product of three terms:
\begin{equation}
\eta_\Apr(m_\Apr)=\eta_{UP} \times \eta_{ECAL}(m_\Apr)\times\eta_{DOWN}(m_\Apr)\; \;,
\end{equation}
where $\eta_{UP}$ is the efficiency for the impinging positron cuts involving upstream detectors, $\eta_{ECAL}$ is the efficiency for the ECAL-based selection, including the missing energy window, and $\eta_{DOWN}$ is that of the downstream VETO and HCAL cuts. This factorization assumes that the three terms are independent; 
this assumption is exact for $\eta_{UP}$, since the upstream track definition is completely independent from the processes occurring in the active target and downstream; on the other hand, the terms $\eta_{ECAL}$ and $\eta_{DOWN}$ are only approximately independent. Indeed, while allowing to evaluate precisely  $\eta_{ECAL}$ accounting for the non-trivial dependence from the LDM parameters (see Sec.~\ref{sec:ECALSignalMC}), the adopted factorization may induce a slightly conservative evaluation of the total signal efficiency $\eta_\Apr$, since it would not take into account all the involved correlations. As an example, events satisfying the ECAL $\chi^2_{shower}$ cut feature a reduced probability to have a high energy deposit in the HCAL, since, by definition, the shower cut selects events with a ``purely electromagnetic'' shower in the ECAL, with no hadron production.  
While $\eta_{UP}$ does not depend on the $\Apr$ mass,
$\eta_{ECAL}$ clearly depends on this parameter, since the residual ECAL electromagnetic shower energy and shape, and thus the energy distribution in the individual detector cells, are different for different $\Apr$ mass values. 
Similarly, the VETO and HCAL cuts efficiency $\eta_{DOWN}$ varies according to $m_\Apr$: the energy deposit in the ECAL depends on the mass of the dark photon, and this result in a different energy leakage probability in the downstream detectors. 

In this analysis, we decided to neglect this dependence, adopting a  data-driven conservative approach to evaluate the combined efficiency of the upstream/downstream cuts, ignoring for the latters the $m_\Apr$ dependency. In summary, the signal efficiency reads:
\begin{equation}
\eta_\Apr(m_\Apr)=\eta_{UP/DOWN} \times \eta_{ECAL}(m_\Apr)\;\;.
\end{equation}


\subsubsection{Upstream/downstream efficiency }\label{sec:UpstreamDownstreamEff}

The combined efficiency of the upstream and downstream cuts $\eta_{UP/DOWN}$ was evaluated by using 70 GeV/c calibration data. As previously discussed, the signal efficiency for the upstream cuts  is the same for any event with a clean impinging 70 GeV/c $e^+$ on the detector, independently on the processes occurring further downstream. 
On the other side, for the VETO and HCAL detectors, using a calibration sample to determine the efficiency results in a conservative estimation, (see Sec.~\ref{sec:downstream_cuts}); however, the adopted procedure has the clear advantage of properly accounting for correlations among the different detector responses and, being totally data-driven, is particularly robust, accounting for any fluctuations in the detectors performance.  The data-set was determined by selecting calibration events and by applying the condition $50\,\, {\rm GeV}< E_{ECAL} < 90\,\, {\rm GeV}$; then, the whole usptream and downstream cut flow was  applied. The overall obtained efficiency was:
\begin{equation}
 \eta_{UP/DOWN} =  (0.6753 \pm 0.0003_{stat} \pm 0.0005_{sys}) \; \;.
\end{equation}
The reported systematic uncertainty was evaluated by repeating the calculation of $\eta_{UP/DOWN}$ slightly varying the ECAL selection, in order to account for the dependence of the result on the quality of the 70 GeV/c $e^+$ sample used for the analysis. 

\subsubsection{ECAL efficiency}\label{sec:ECALSignalMC}

The efficiency for the ECAL cuts on signal events was computed through a methodology involving both MC simulations and data. We decided to adopt, wherever possible, a data-driven approach. Observables like the $\chi^2_{shower}$ are highly sensitive to the low-energy tails of the developing electromagnetic shower measured by the peripheral cells, making a reliable description of these variables using MC simulation very challenging. For all the ECAL cuts described in Sec~\ref{sec:downstream_cuts}, except the pre-shower and missing-energy cut, we evaluated the  signal efficiency using the same di-muon event sample used for the $\chi^2_{shower}$ cut optimization,  exploiting the kinematic signature that these share with the expected signal. The di-muon events were selected with the upstream detectors, to identify a clean impinging 70 GeV/c track, and by requiring a MIP-like energy deposit in the three downstream HCAL modules. As a further selection, events with ECAL ``out-of-time'' energy fraction greater than $50\%$ were rejected, in order to clean the sample from events affected by pile-up. These corresponded to a fraction $\eta_{pile-up}= 0.980\pm0.005_{sys}$, with the uncertainty evaluated by slightly varying the ``out-of-time'' energy fraction defining the selection\footnote{This effect was accounted for in the analysis by correcting the effective number of \pot of the measurement.}.
The event sample was then processed applying the cuts, 
in order to assess the corresponding efficiency. The obtained value is $\eta_{ECAL\,topology} = (0.946 \pm 0.007_{sys})$, where the systematic uncertainty was evaluated by repeating the calculation using a slightly modified condition to isolate the di-muon event sample. 
In light of the adopted procedure, the resulting efficiency $\eta_{ECAL\,topology}$ was multiplied by an additional $m_\Apr$-dependent factor $C_{\chi^{2}\,MC}$, obtained with the MC-based procedure described in Appendix~\ref{app:chi2_correction};  this correction 
accounts for the minor differences in the ECAL shower development for di-muon and  resonant-annihilation signal events, resulting in a slightly reduced efficiency of the $\chi^2_{shower}$ cut for signal compared to di-muon events.

\begin{figure}[t]
    \centering
    \includegraphics[width=0.45\linewidth]{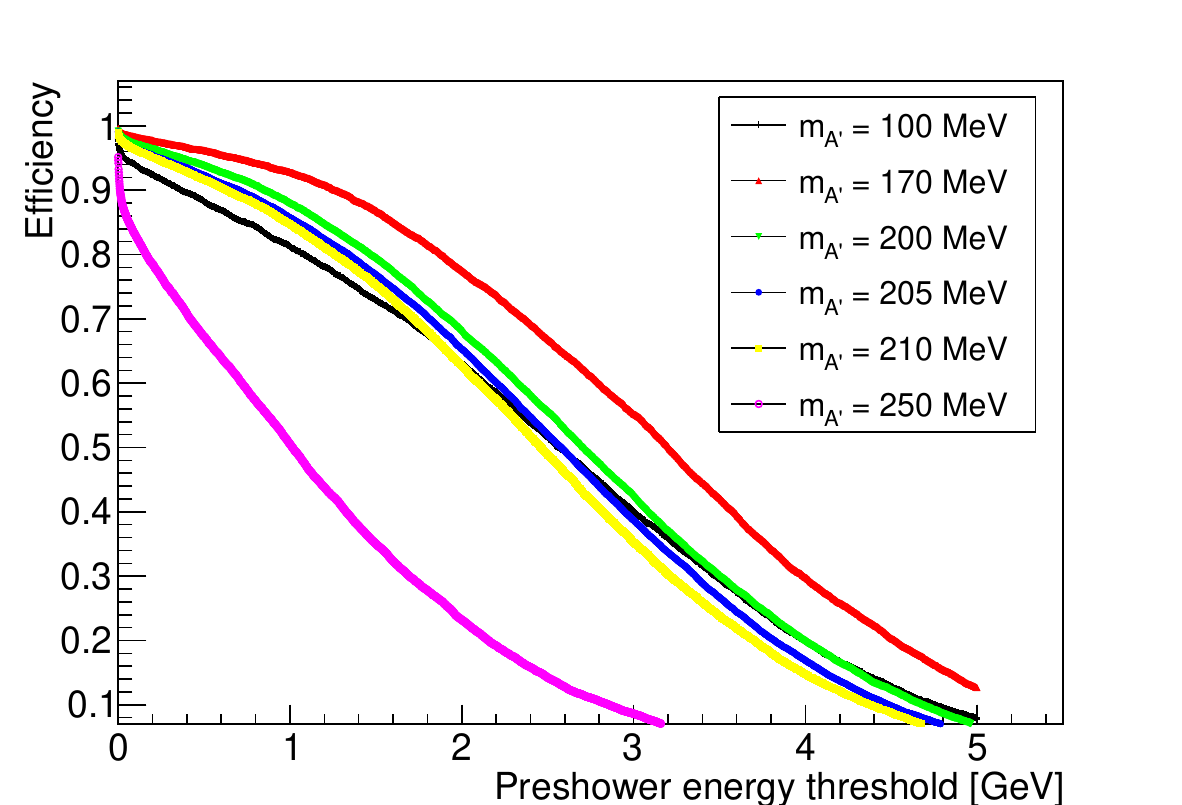}
    \includegraphics[width=0.45\linewidth]{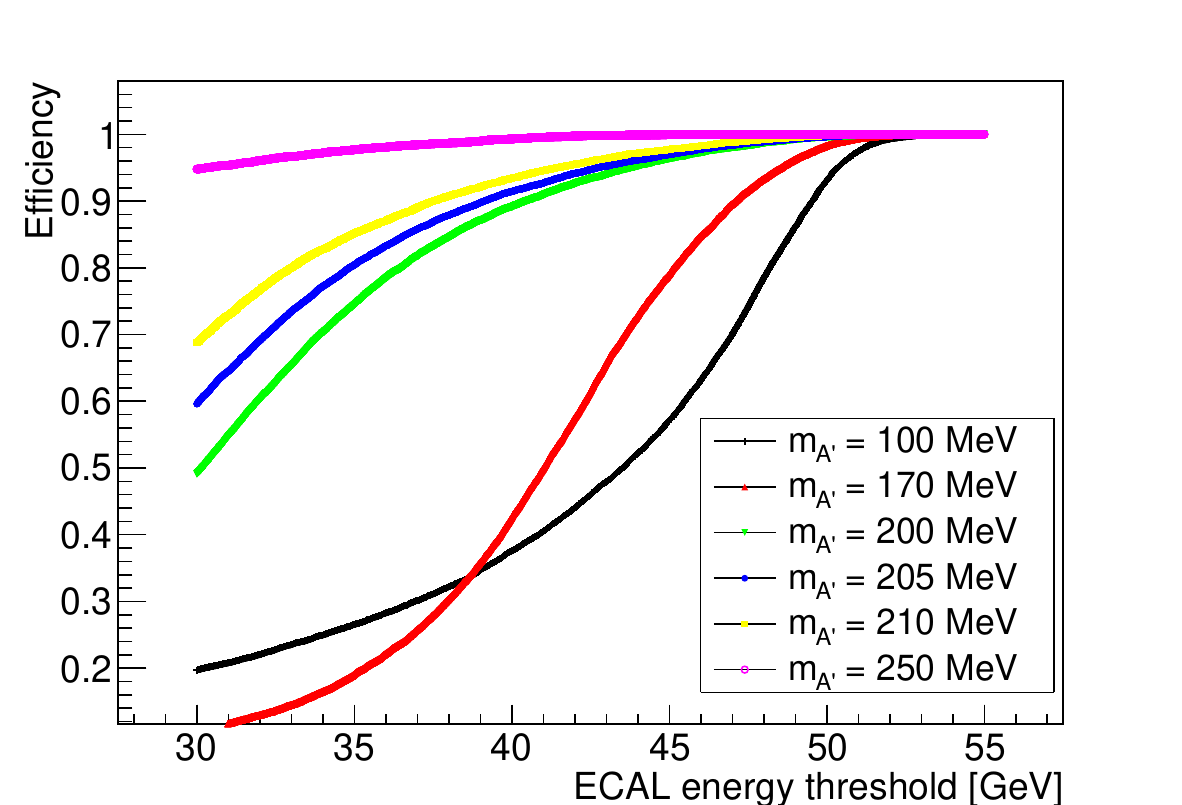}
    \caption{Left (right) panel: ECAL efficiency $\eta_{ECAL}(m_\Apr)$ for $\Apr$ signal events from $e^+e^-$ annihilation as a function of the pre-shower (ECAL) energy threshold, for different dark photon masses. Both the plots refer to the case $\alpha_D=0.1$.}
    \label{fig:ECALeff}
\end{figure}

The effect of the pre-shower and the missing-energy cuts on signal efficiency, instead, was evaluated via MC simulations. The generation of a MC sample of $\Apr$ events was performed through a full Geant4 simulation of the NA64$e$ setup based on the DMG4 package~\cite{Oberhauser:2024ozf,Bondi:2021nfp}. We simulated several samples, varying  m$_\Apr$ in the range $1$ MeV - $10$ GeV. For each $\Apr$ mass under consideration, the simulation was performed with the nominal choice of the kinetic mixing $\varepsilon_0=10^{-4}$, two benchmark cases of $\alpha_D=0.1$ and $\alpha_D=0.5$, and the common choice of $m_\Apr=3m_\chi$; both resonant annihilation and $\Apr$-strahlung processes were simulated. To optimize the computation time, an ad-hoc bias scheme was implemented,
involving a cut on the $\Apr$ production cross section below a certain electron/positron energy, as well as a scaling factor applied for larger values (see Ref.~\cite{Andreev:2021fzd} for a complete description). Furthermore, we adopted an improved simulation scheme, exploiting the Geant4 \texttt{StackingAction} system to run the full ECAL shower simulation only for events in which the $\Apr$ production takes place, resulting to a factor $\times 100$ improvement in computation time - further details are provided in Appendix~\ref{app:stackingAction}. A summary of the obtained results is reported in Fig.~\ref{fig:ECALeff}, showing for different values of $m_\Apr$ the signal efficiency for $\Apr$ particles produced by $e^+e^-$ annihilation as a function of the pre-shower and ECAL thresholds, in the ranges relevant for the analysis. For the latter, the larger variation is observed for $m_\Apr=170$~MeV, corresponding to a positron resonant energy of 28.28 GeV, and thus to an ECAL deposited energy of 41.72 GeV - for this mass value, a small change of the ECAL energy threshold would result in a large variation of the $\Apr$ yield within the signal region.

\subsection{Background sources}\label{sec:backgrounds}

In this Section, we provide an estimate of the expected number of background events mimicking the signal signature, i.e. events induced by SM processes satisfying all the analysis cuts described in Sec.~\ref{sec:cuts}.
The estimation of the expected backgrounds at the level of a single event sensitivity for the accumulated statistics of $1.596 
\times 10^{10}$ \pot via a full MC simulation would require a very large computing time. Furthermore, a ``brute-force'' strategy would entirely rely on the implementation within the simulation software of all background processes, with an accuracy level capable of predicting the yield of signal-mimicking rare-events at this level. Therefore, our estimates are based on a combination of MC simulations, and
direct measurement with the beam. When possible, we tried to extrapolate the expected contribution of a given background source directly from the measured data. 
The dominant background sources can be divided in three general categories, (a) the interaction of the impinging positron beam with upstream beam line materials, (b) the decay of contaminating muons and hadrons in a final state involving a soft electron plus neutrinos, and (c) the energy leakage due to penetrating particles produced in the active target, exiting from it undetected. A summary of each contribution, as well as the final estimate of the overall yield, is reported in Tab.~\ref{tab:bck}. 

\begin{table}[t] 
\begin{center}
\vspace{0.15cm}
\begin{tabular}{lr}
\hline
\hline
Background source& Background, $n_b$\\
\hline
(i) $e^+$ hadronic interactions in the beam line &  $0.08 \pm 0.07_{stat} \pm 0.03_{sys} $ \\
(ii) $\mu$ decays & $(6 \pm 1)\times 10^{-4}$ \\
(iii) $\pi,~K$ decays& $(9\pm2)\times 10^{-5}$, $(1.5 \pm 0.6) \times 10^{-4}$ \\
(iv) Hadrons interactions in the target & $(2.2\pm 0.6)\times 10^{-5}$ \\
(v) Di-muons leakage &  $< 0.8 \times 10^{-5}$ \\
\hline 
Total $n_b$ (conservatively)   & $0.08 \pm (0.07)_{stat} \pm (0.03)_{sys}$ \\ 
\hline
\hline 
\end{tabular}
\end{center}
\caption{\label{tab:bck} Summary of the expected backgrounds contribution in the measurement, for a total accumulated statistics of $1.596\times 10^{10}$ \pot.}
\end{table}

\subsubsection{Upstream interactions} \label{sec:upstreamInteractions}

The dominant background source for the measurement is the interaction of the primary 70 GeV/c positron beam with upstream beam line components (such as e.g. the entrance and exit windows of the vacuum chamber, the residual gas in the latter, the SC, MM, and GEM materials, the Straw detectors, see Fig.~\ref{fig:NA642023setup}), occurring either directly or via the radiative emission of a intermediate Bremstrahlung photon. 
This may give rise to a background event if the secondary hadrons falls out of the acceptance of the detector, while the soft final state $e^+$ impinges on the ECAL giving rise to a low-energy electromagnetic shower, with no associated activity in the VETO and HCAL detectors. In order to suppress this background, the NA64 detector assembly was recently optimized to have one of the hadronic calorimeter modules (HCAL3) installed at zero-angle downstream, to detect neutral secondaries not deflected by the dipole magnet, and installing a prototype version of the so-called ``veto hadronic calorimeter'' (VHCAL) between the dipole magnet and the ECAL to further enhance the angular detection acceptance for charged secondaries deflected along the positron trajectory. Still, the lower positron beam energy, compared to the nominal 100 GeV/c electron configuration, results in a larger emission angle for secondaries, and thus to a higher background contribution.

\begin{figure}[t]
    \centering
    \includegraphics[width=0.48\linewidth]{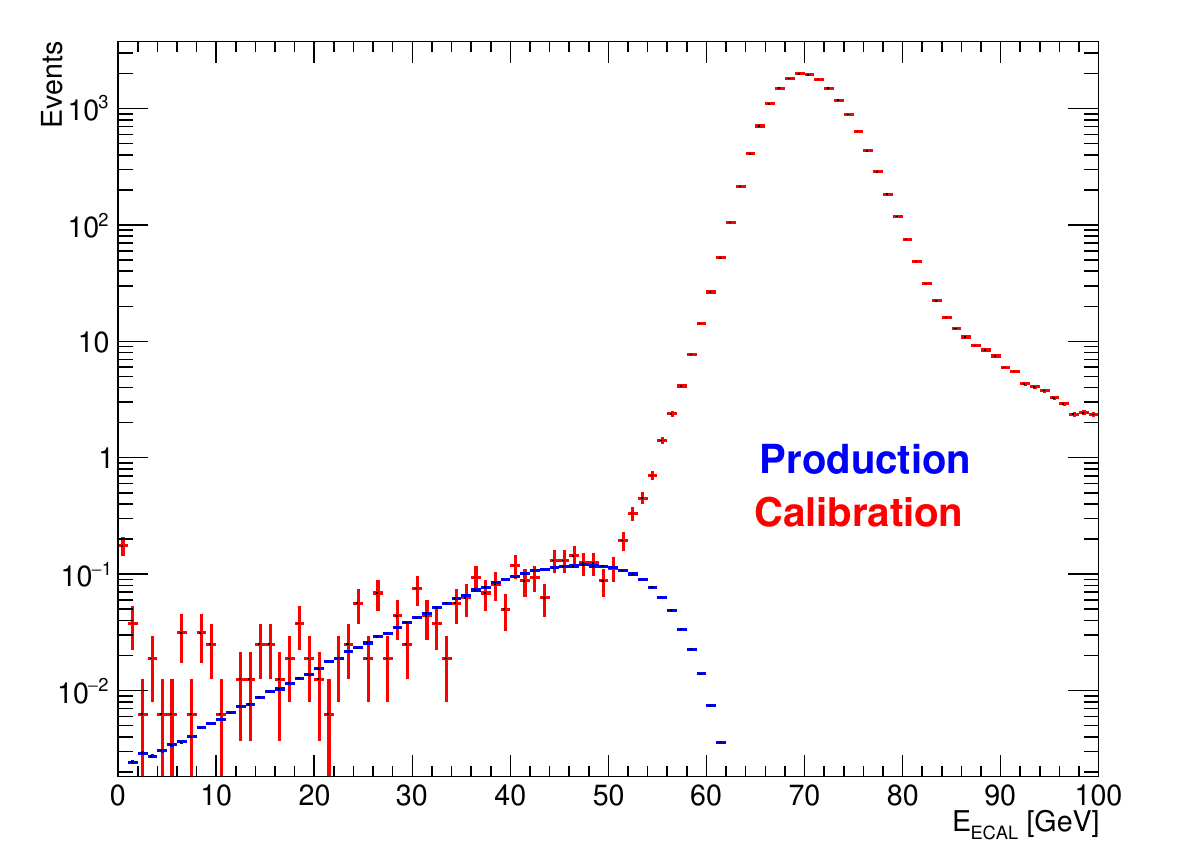}  
    \includegraphics[width=.43\linewidth]{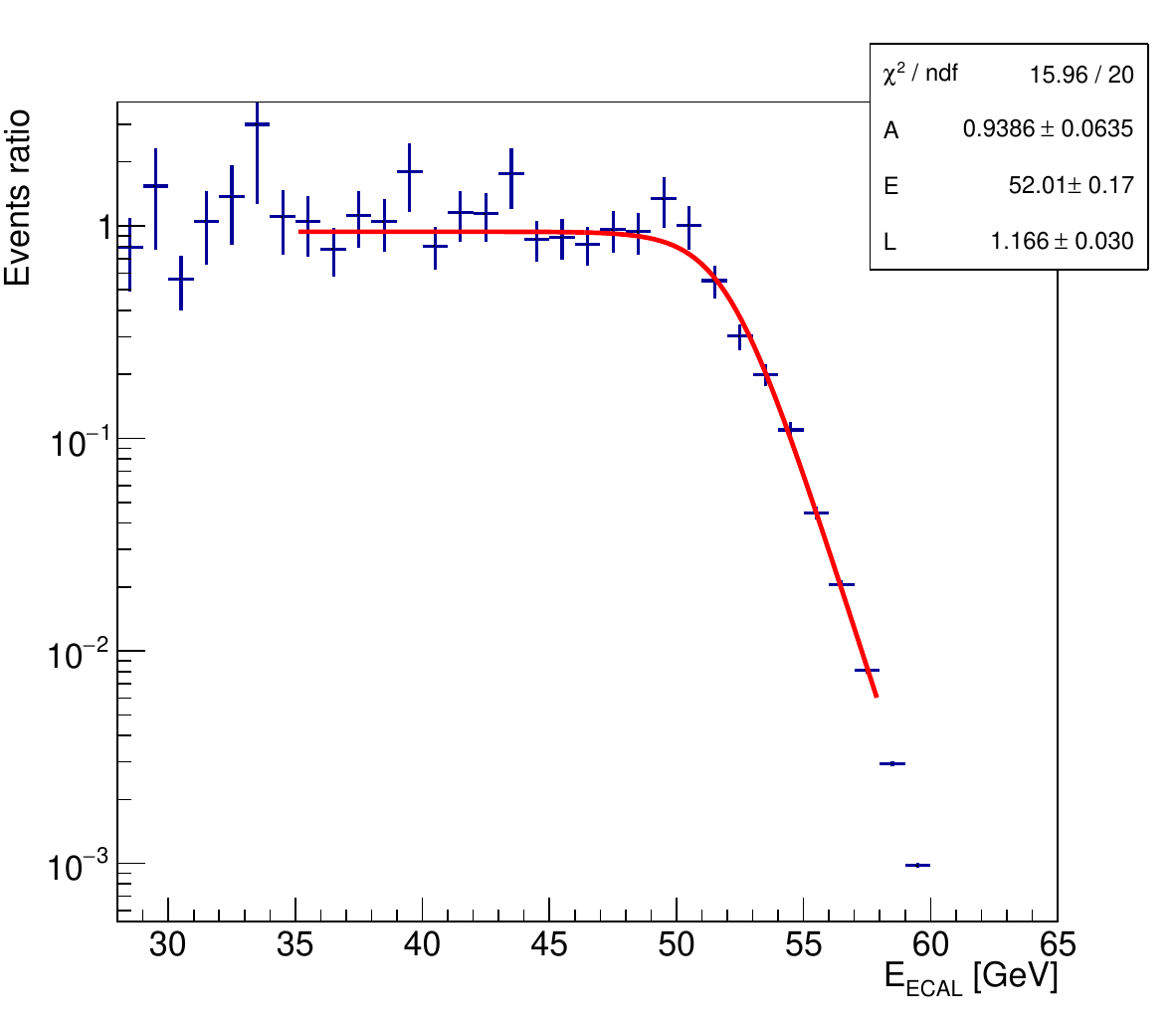}
    \includegraphics[width=0.48\linewidth]{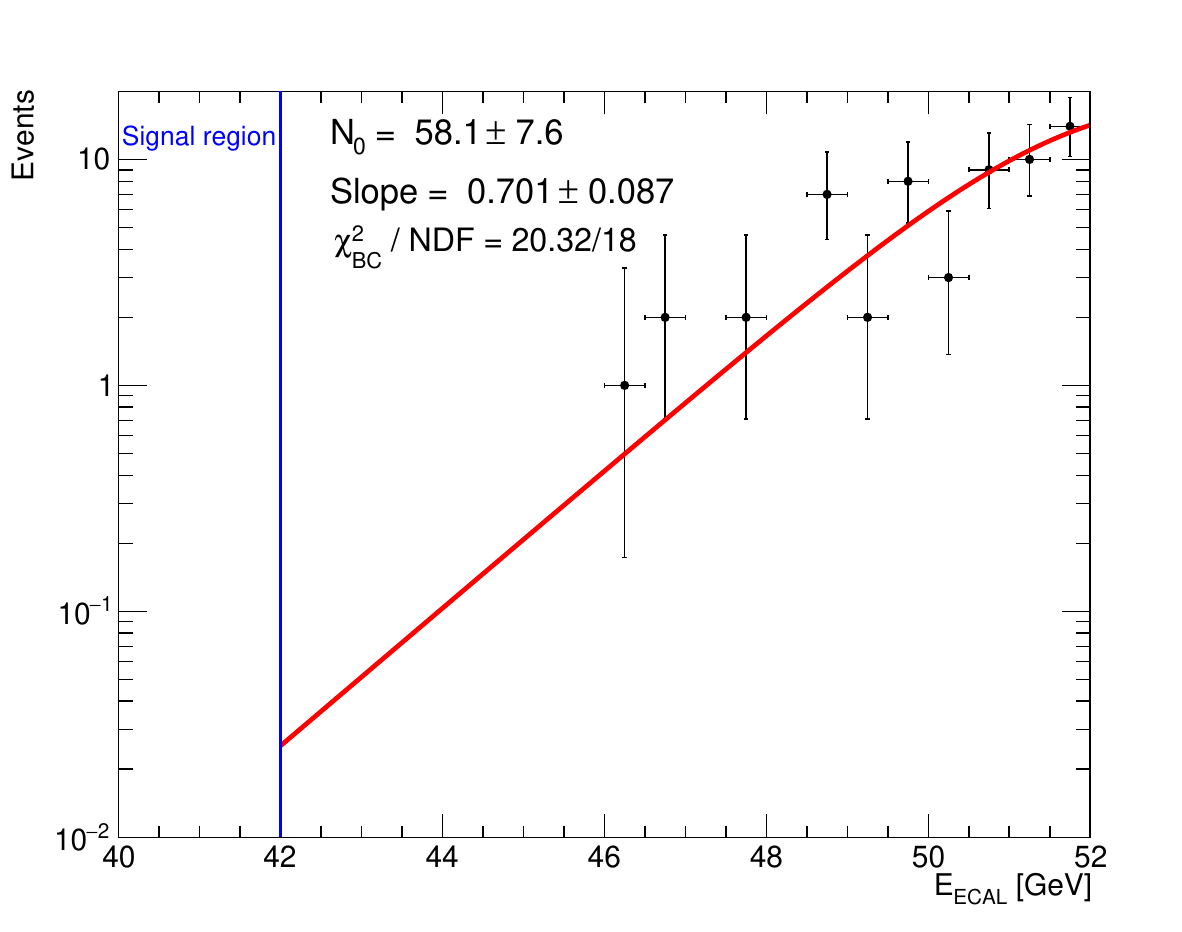}
    \caption{Top-left panel: the ECAL energy spectrum for 70 GeV/c positrons selected by the upstream cuts in a production run (blue) and in a calibration run (red), both scaled to unity in the 20-40 GeV range. Top-right panel: the ratio of the two distributions, together with an unbinned maximum likelihood fit performed with a sigmoidal function to extract the value of the ECAL missing energy trigger threshold $E_{thr}$. Lower panel: the ECAL energy distribution of all events selected after applying all the analysis cuts, together with the result of an extended maximum-likelihood fit performed with an exponential function multiplied by a sigmoidal to evaluate the expected background yield due to upstream interactions. We underline that in the 42-46 GeV range no events have been observed. See text for further details.}
    \label{fig:enh}
\end{figure}

To estimate the corresponding yield, we adopted a fully data-driven approach, extrapolating it from the events in a ``sideband'' control region, defined by the condition $E_{ECAL}>42$~GeV, $E_{HCAL}<1$~GeV, using an exponential distribution. Specifically, we considered all the events in the production runs and applied to them all the selection cuts, while still keeping the signal region blinded. The corresponding ECAL energy distribution is reported in Fig.~\ref{fig:enh}, bottom panel. It shows a decreasing monotonic trend for low $E_{ECAL}$ values, partially affected by the ECAL missing energy threshold applied at the trigger level, $E_{thr}$, for $E_{ECAL} \simeq 52$~GeV. 

Given that the extrapolation result is strongly affected by the dominant event yield in this energy range, a proper description of the threshold was necessary. We decided to parameterize the corresponding effect to the ECAL energy spectrum through a sigmoidal function\footnote{We use the sigmoidal parameterization $S(x)=\frac{1}{1+e^{-x}}$.} $S(\frac{E_{thr}-E_{ECAL}}{\lambda})$ multiplying the exponential term, in which the parameter $\lambda$ describes the steepness of the effective threshold rise. We determined both $E_{thr}$ and $\lambda$ from data, exploiting calibration events collected during the and comparing the corresponding ECAL energy spectrum to that of a production run, in the region close to the 50 GeV trigger threshold. From both datasets, we selected events applying the upstream cuts to isolate well identified 70 GeV/c impinging positrons, and we also required the energy deposited in the pre-shower to be larger than 350 MeV. We rejected pile-up events, possibly distorting the energy spectrum, by applying a cut on the total out-of-time energy in the central ECAL cell. The obtained spectra, reported in Fig.~\ref{fig:enh}, top-left panel, were both normalized to unity in the 20 GeV - 40 GeV range, and finally the production-run distribution was divided by the calibration trigger one. The obtained result is shown in Fig.~\ref{fig:enh}, top-right panel, together with the result of a fit performed with the sigmoidal function  $A\times S(\frac{E_{thr}-E_{ECAL}}{\lambda})$, with $A$, $\lambda$ and $E_{thr}$ free parameters. The obtained trigger threshold value is $E_{thr}=(52.0\pm 0.2)$~GeV. The stability of the threshold was evaluated by repeating this study for different production runs, and found to be within 0.2$\%$.

To extrapolate the background yield due to upstream interactions from the sideband region, we adopted an extended unbinned likelihood approach exploiting the RooFit package~\cite{Verkerke:2003ir}, considering the events in the ECAL energy range between 42 GeV and 52 GeV. We adopted the PDF $f(E_{ECAL})=N_0 \times f_{norm}\times exp(p_{slope}\, E_{ECAL})\times S(\frac{E_{thr}-E_{ECAL}}{\lambda})$, i.e. an exponential corrected for the trigger threshold effect as previously discussed, where $N_0$ and $p_{slope}$ are free parameters. For the sigmoidal, we fixed $E_{thr}$ and $\lambda$ to the previously determined values. Finally, $f_{norm}$ is the function integral in the aforementioned energy range for $N_0=1$. The fit was performed using the RooFit package~\cite{Verkerke:2003ir}. The PDF was then integrated within the signal region to determine the final yield, and the standard RooFit method was adopted to evaluate the effect of the statistical uncertainty on the obtained fit parameters. To check the fit robustness, we evaluated the Baker-Cousin~\cite{Baker:1983tu} $\chi^2$ value considering a binned representation of data, obtaining the result $\chi^2_{BC}/NDF=20.32/18$. The systematic uncertainty for the background prediction was evaluated by repeating the fit procedure with different intervals, as well as varying $E_{thr}$ and $\lambda$ within their uncertainties. The obtained result for the expected number of backgrounds due to the $e^+$ hadronic interactions in the beam line was $(0.08 \pm 0.07_{stat} \pm 0.03_{sys})$ events. We also checked the effect of a different PDF parametrization. We tried to fit the distribution with a Crystal Ball function, a pure exponential function with no sigmoidal correction, and a power-law function multiplied by the sigmoidal. The Crystal Ball function did not provide a satisfactory fit result and was thus neglected. Instead, the power-law function and the pure exponential function resulted in a variation of the background yield lower than the variation due to the original function fit range modification.  
Finally, the systematic uncertainty due to the assumption that the data distribution follows the  exponential parametrization up to $E_{ECAL}=0$ GeV was estimated from Monte Carlo simulations. We simulated events due to the upstream interactions of impinging positrons with the beam line materials in the current setup, for a total MC statistics of $5.7\times10^9$ \pot (see Appendix~\ref{app:upstremInteractionsSimuDataComparison} for further details). We applied to MC events the same analysis cuts employed for the data, apart from the VHCAL one -- indeed, if the latter is employed, no events at all are observed in the signal region. The result is shown in Fig.~\ref{fig:MCcomparison}; on this, an exponential function fit was performed. We compared the integrated number of events in the $E_{ECAL}$ range between 0 and 42 GeV between the original MC sample and the exponential parameterization, obtaining respectively $N_{MC}=(1.57\pm0.07)\times 10^{-8}/$\pot and $N_{expo}=(1.51\pm0.07)\times 10^{-8}/$\pot. This $\simeq4\%$ difference was accounted as an additional systematic uncertainty for the predicted number of upstream interactions events.

\subsubsection{In-flight particles decay}\label{sec:InFlightDecay}

The in-flight decay of a beam-contaminating hadron or muon to a final state involving a soft electron and one or more energetic neutrinos can mimic the signal signature, if the impinging particle is identified as a positron. This may happen if the hadron/muon interacts with the upstream beam line elements (in particular, the downstream vacuum window), emitting a $\delta$ electron that hits the SRD, giving rise to a signal similar to the one expected from a 70 GeV/c $e^+$\footnote{A sub-dominant effect creating false-positive SRD signals for impinging hadrons/muons is the random in-time superposition with a low positron from the beam tail within the typical response time of the detector ($\simeq 30\div40$ ns), if the $e^+$ emits enough synchrotron radiation to satisfy the SRD cut and then is deflected away by the bending magnet. This effect was found to be approximately one order of magnitude smaller than that due to the emission of a $\delta$-ray, and will thus be ignored in the following.}. We measured the probability for this false-positive scenario $p_\delta$ exploiting data collected during a so-called ``hadron calibration'' run. 
In this configuration, the H4 beam line is operated without the photon 4-mm thick lead converter after the XTAX aperture~\cite{Andreev:2023xmj}. Particles entering in the beam line and further transported downstream after septum-magnet selection are mostly protons and $\pi^+$ produced by the decay of a $\Lambda$ or $K^0$ particle, arising in the T2 target and then propagating straight downstream. To measure $p_\delta$ from this dataset, we isolated impinging hadrons considering events with an energy deposit in the ECAL compatible with that expected for a MIP - $E_{ECAL} \simeq 300$ MeV -
 and full energy deposit in the HCAL. We then evaluated the fraction of the events satisfying the SRD cut, for different values of the corresponding energy threshold. We observed a strong dependency of the result on the latter parameter, with $p_\delta$ ranging from $(3.9\pm0.1)\times 10^{-2}$ for 0.5 MeV threshold to $(8\pm3)\times 10^{-5}$ for 2.5 MeV threshold. 

\begin{figure}[t]
    \centering
    \includegraphics[width=0.45\linewidth]{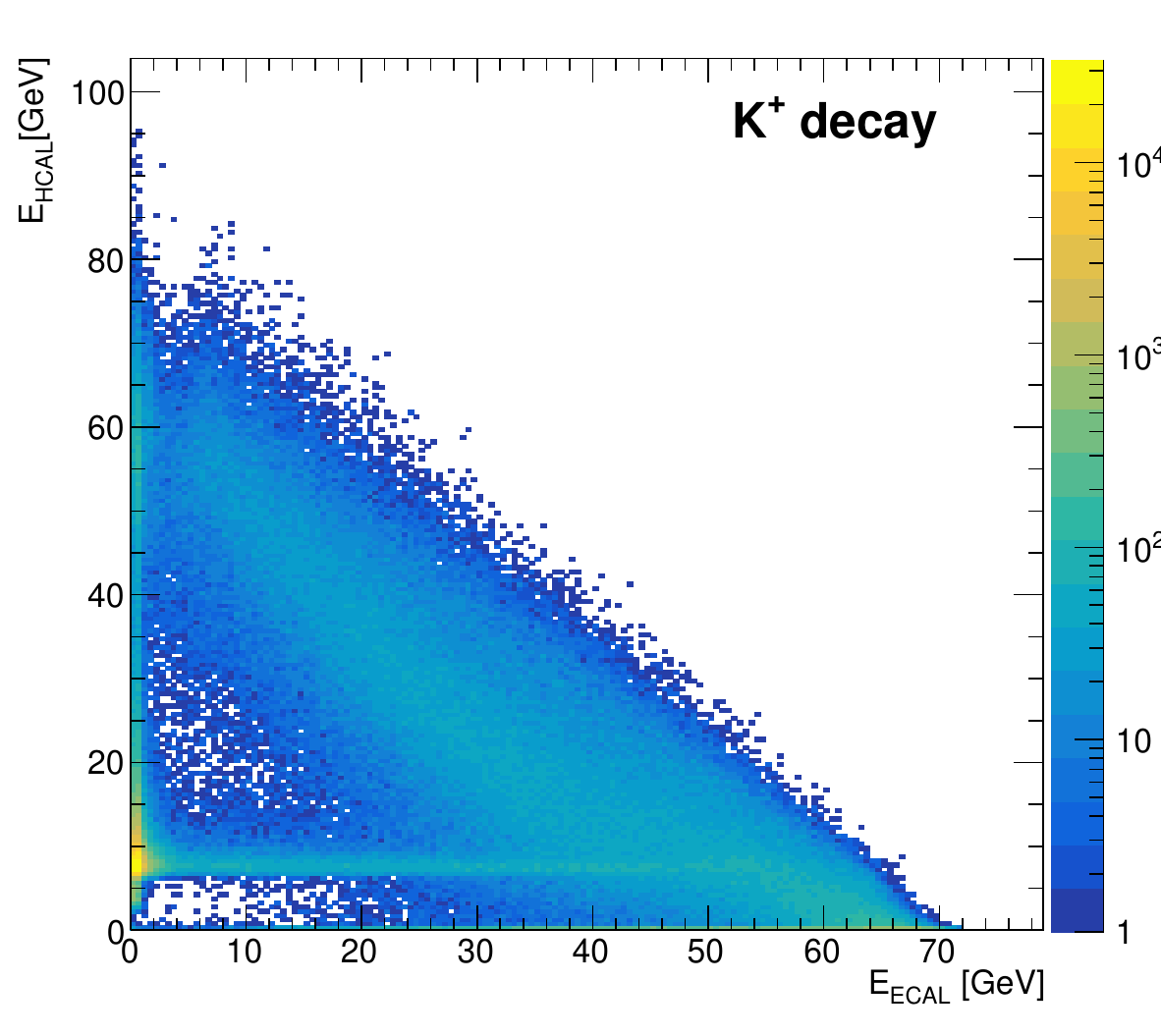}     \includegraphics[width=0.45\linewidth]{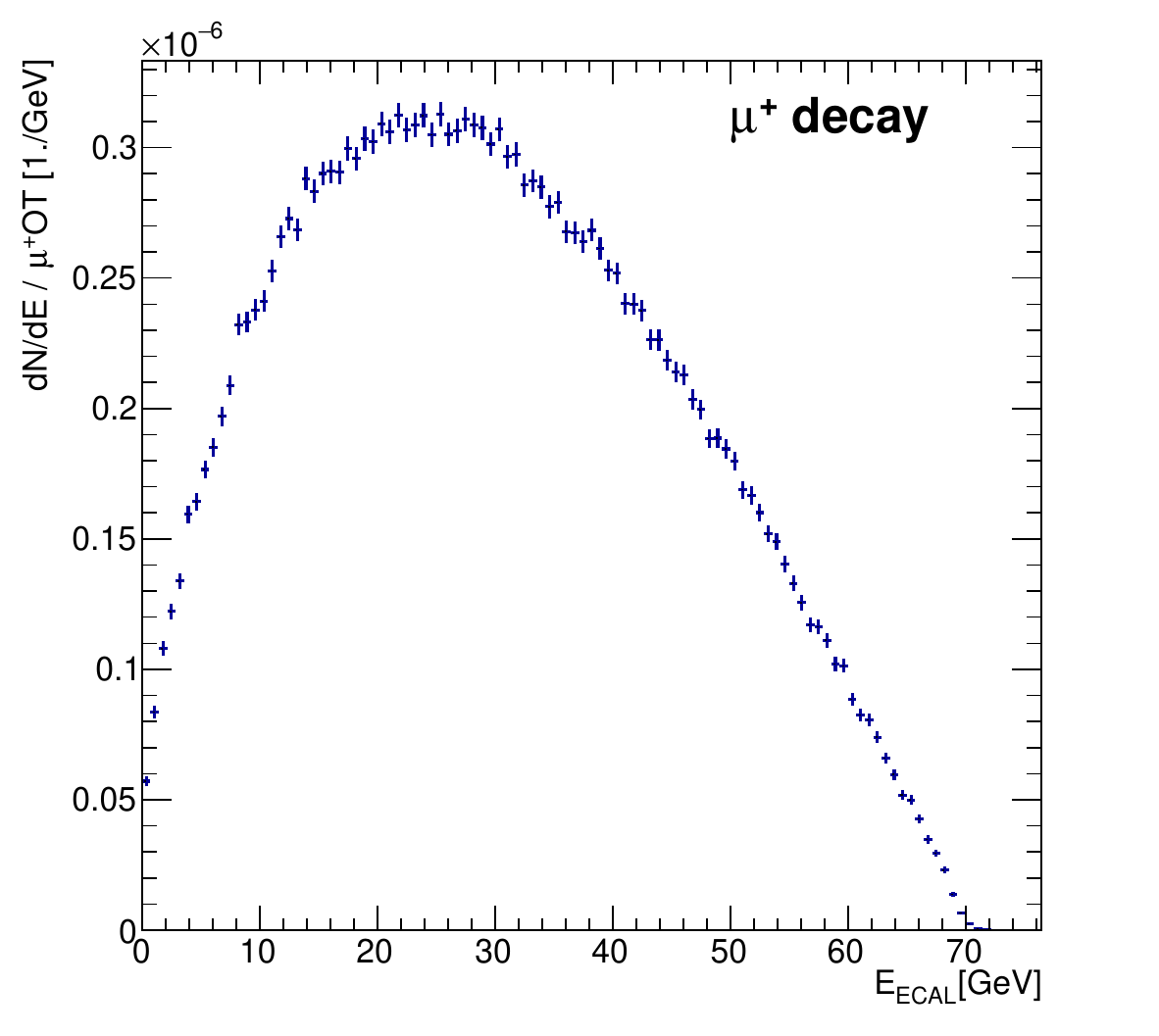}
    \caption{Left: HCAL vs ECAL energy distribution for $K^+$ in-flight decay events. Right: ECAL energy distribution for $\mu^+$ in-flight decay events, normalized to single impinging muon.}
    \label{fig:bckInFlight}
\end{figure}

 The dominant particle in-flight decay background arises from the decay of a misidentified kaon to the final state with one positron, one $\pi^0$, and one neutrino ($K^+ \rightarrow e^+ \pi^0 \nu_e$ -- so called $K_{e3}$ decay), when the  $\nu_e$ carries away a large fraction of the beam energy and the positron and the two photons from $\pi^0$ decay are collimated forward and give rise to a single, low energy EM shower in the ECAL. The contribution of the $K^+ \rightarrow e^+ \nu$ decay is strongly suppressed by the corresponding branching ratio, $\simeq 1.6 \times 10^{-5}$. Similarly, the $K^+\rightarrow \mu^+ \nu$ decay can mimic the signal if the muon is poorly detected or decays in-flight, while the $K^+ \rightarrow \pi^0 \pi^+$ can do so if the $\pi^+$ is poorly measured, or interacts hadronically in the ECAL giving rise to a missed neutral final state. The overall background yield was estimated computing the probability for an impinging kaon to decay within the NA64$e$ detector acceptance, mimicking the signal signature, and multiplying it for the overall misidentified kaons yield. When the H4 beam line is operated in high-purity 70 GeV/c $e^+$ mode, kaon contaminants result mainly from the conversion of a high-energy photon from the T2 target to a $K^+/K^-$ pair, or from the direct off-axis emission of a kaon in the T2 target, deflected back toward the XTAX hole by the sweeping magnets. A rough estimate of the $K^+$ contamination at 70 GeV/c \textit{just after the converter} obtained from MC reads $\eta_k\sim1.6\times10^{-6}$~\cite{Andreev:2023xmj}. This is further suppressed by kaon decays happening in-flight along the $\simeq 540$-m long H4 beam line, resulting to $\eta_k\simeq 6.8\times10^{-7}$ at the NA64 detector location. It follows that the total number of impinging $K^+$ corresponding to the accumulated statistics reads approximately $N_K \simeq 1.1 \times 10^{4}$, with a sizable systematic uncertainty of about $40\%$ associated to $\eta_k$. The probability for these impinging kaons 
 to give rise to backgrounds was computed through a FLUKA-based MC simulation~\cite{Ahdida:2022gjl,Battistoni:2015epi,Vlachoudis:2009qga} of the NA64$e$ setup, applying to the generated events all the upstream selection cuts, apart from the SRD one, and then evaluating the fraction satisfying the signal condition $E_{ECAL}<42$~GeV, $E_{HCAL}<1$~GeV. The HCAL vs ECAL energy distribution for events in which the primary $K^+$ decays before impinging on the ECAL is shown in Fig.~\ref{fig:bckInFlight}, left panel. $K_{e3}$ decays manifest as an accumulation of events in the signal-like region, $E_{HCAL}\approx$ 0, with probability per impinging kaon $I_K=(7.55 \pm 0.09) \times 10^{-5}$. In total, accounting for the efficiency of the ECAL, VETO, and HCAL signal selection cuts, not applied to the MC sample, the expected background yield is $B_K=(1.5 \pm 0.6) \times 10^{-4}$.

In-flight $\pi^+$ decay contribute to the background yield mainly from the $\pi^+ \rightarrow e^+ \nu_e$ channel, although this is suppressed both by the low branching ratio (BR$=1.23\times10^{-4}$) and by the large $\pi^+$ mean life, resulting to a boosted decay length of about $L_d=\gamma c\tau=3.9$ km at 70 GeV. The relative fraction of $\pi^+$ in the hadronic contaminants \textit{at the T2 target} for the H4 beam was also computed via MC and found to be $f_\pi \simeq 7\%$~~\cite{Andreev:2023xmj}, with a $\simeq 10\%$ uncertainty, i.e. $f_\pi=(7.0\pm 0.7)\%$. Accounting for the suppression due to in-flight decay in the $\sim$500-m long H4 beam line before the detector and considering the intrinsic hadronic contamination of the beam $\eta_h=(0.457\pm0.007)\%$, the total number of impinging $\pi^+$ corresponding to the whole positron measurement was found to be about $N_{\pi^+}= (4.4\pm 0.4)\times 10^{6}$. Similarly as before, the probability for an impinging $\pi^+$ to decay and mimic the signal signature was evaluated from MC and found to be $I_\pi=(1.2 \pm 0.1)\times 10^{-7}$, resulting to an overall background prediction of $B_\pi=(9 \pm 2) \times 10^{-5}$.

Finally, the in-flight decay of an impinging muon to the $e^+\nu_e$ final state involving a soft electron and an energetic neutrino can mimic the signal signature. This background is strongly suppressed by the large lifetime of the muon, $\tau \approx 2.2\,\mu s$, resulting to a mean life in the laboratory frame $L_d=\gamma c \tau \simeq 440$~km at 70 GeV. This suggests a decay probability within the acceptance of the NA64$e$ detector, in the region between the MBPL magnets and the ECAL, of about $L/L_d\simeq 3.5\times 10^{-5}$, where $L\simeq 15$~m is the useful length for the decay; this number still has to be multiplied by the probability for the final state electron energy to be lower than $\simeq 42$~GeV, of about $50\%$. As before, we consolidated this estimate by running a dedicated FLUKA simulation generating 70 GeV/c $\mu^+$ upstream the NA64$e$ detector. The measured ECAL energy spectrum for events in which the muon actually decayed upstream the ECAL, and no activity in the HCAL is present, is shown in Fig.~\ref{fig:bckInFlight}, right panel, normalized to the total number of simulated particles. The number of events with an energy deposit $\lesssim 42$~GeV is $I_\mu=1.1 \times 10^{-5}$, in reasonable agreement with the previous estimate. The total number of impinging $\mu^+$ for the whole run was computed from the measured value of the muons fraction in the H4 high-purity $e^+$ beam~\cite{Andreev:2023xmj}, $\eta_\mu\simeq 2\times 10^{-5}$, resulting in $N_\mu=\eta_\mu \times N_{h+e}\simeq 3.2\times 10^{5}$, with an uncertainty of about $10\%$. Including the efficiency for VETO, HCAL, and ECAL cuts, the overall background was found to be $B_\mu=(6 \pm 1)\times 10^{-4}$.

\subsubsection{Leakage effects}

Penetrating particles such as muons or neutral hadrons produced in the ECAL by the interaction of an impinging 70 GeV/c positron or misidentified hadron can give rise to a background event if they are not properly identified and tagged by the downstream detector elements - the VETO and the HCAL. 

For misidentified impinging hadrons, the most critical contribution comes from final states involving neutral particles only, since no suppression from the high-efficiency VETO detector occurs. We evaluated the corresponding yield starting from the hadron calibration dataset, with a statistics significantly larger than the overall number of contaminating hadrons for production runs $N^0_h \simeq (7.3\pm 0.1)\times 10^{7}$, out of which $N_h=N^0_h\times p_\delta=(1.8 \pm 0.3 )\times 10^{4}$ are expected to be misidentified as positrons by the SRD cut, for the nominal 2 MeV threshold value.  First, we computed the fraction $f_n$ of events with a hard interaction in the ECAL with only neutral-secondaries exiting from the ECAL, by applying all the upstream selection cuts, apart from the SRD one, that on the pre-shower energy, to mimic the trigger condition, and the VETO one, obtaining $f_n=(0.44 \pm 0.02)$ -- incidentally, we observe that all the selected events satisfy the energy-conservation condition $E_{ECAL}+E_{HCAL}=E_0$ within the experimental resolution of the two detectors. Next, we computed the punch-through probability for a single HCAL module, applying again all the upstream cuts but the SRD one, and requiring an energy deposit in the ECAL compatible with a MIP signature. The probability $f_{punch-through}$ was obtained as the ratio among the number of events with a MIP-like signature in the HCAL0 detector and a full-energy deposit in HCAL1 with respect to the number of events with full energy deposit in either HCAL0 or HCAL1; the obtained value was $f_{punch-through}=(0.14\%\pm0.02\%)$. Finally, we measured the fraction $f_h$ of overall impinging hadron events lying on the energy-conserving band and satisfying $E_{ECAL}<42$~GeV, obtaining $f_h=(0.44\pm0.02)$. Assuming, very conservatively, that all events with neutral-only secondaries actually produce a single, leading secondary hadron exiting from the ECAL, the expected background yield can be estimated as $B_{h}^{punch-through} = N_h \times f_h \times (f_{punch-through})^3 = (2.2\pm 0.6)\times 10^{-5}$, where $(f_{punch-through})^3$ is justified by the presence of three subsequent HCAL modules.

\begin{figure}[t]
    \centering
    \includegraphics[width=1.01\linewidth]{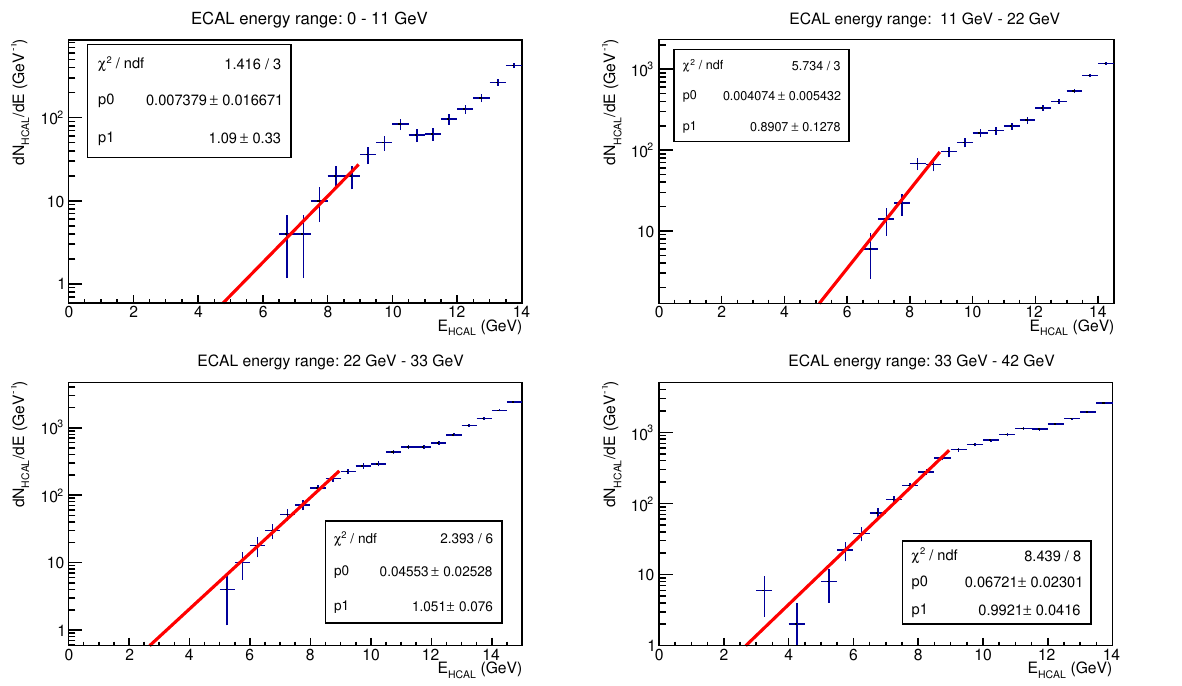}
    \caption{Di-muon events HCAL energy distribution distributions for different ECAL energy intervals, together with the result of the fit with an exponential curve used to extrapolate the yield in the low-energy region.}
    \label{fig:dimuBck}
\end{figure}

For an impinging positron, instead, due to the significantly suppressed production of secondary hadrons, the dominant source of potential leakage is the production of di-muon pairs, if both particles are not detected by the VETO and the HCAL, and only the residual energy deposit from the EM shower is measured in the calorimeter. To estimate this background, first we proceeded to measure the number of di-muon events with an energy deposit in the ECAL in the 0 GeV - 42 GeV range that are reconstructed with an HCAL energy lower than 1 GeV. These events can be caused by a poor energy reconstruction of the di-muon pair in the HCAL, and can produce a background if they are not rejected by the VETO cut. For this study, we applied to the data the di-muons selection criteria described in Sec.~\ref{sec:dimuonsDataMC}, modifying those on the HCAL to include only the upper threshold, to not bias the extrapolation to the signal region. The selected sample was divided into four subsets, based on the ECAL energy, to take into account any differences depending on the di-muon pair energy, and for each the differential distribution as a function of the HCAL energy $\frac{dN_i}{dE_{HCAL}}$ was built (see Fig.~\ref{fig:dimuBck}). We then fitted the low energy part of the spectra with an exponential function, and we used it to extrapolate the expected number of events for each bin in the signal region, defined by the condition $E_{HCAL}<1$~GeV. The four obtained values were summed, so to get the expected number of events in the signal box $N^{2\mu}_{SB}$. The systematic uncertainty was evaluated by varying the upper limit of the energy interval for the exponential fit by $\pm 1.5$ GeV. The background yield was finally obtained by multiplying $N^{2\mu}_{SB}$ by the VETO inefficiency for di-muon events, found to be $(1.5\pm1.0)\times 10^{-5}$. In conclusion, the expected background contribution is $N^{2\mu}_{BKG} \leq N^{2\mu}_{SB \,\,max} \times 1.5 \times 10^{-5} < 0.8 \times 10^{-5}$.

\subsection{``Di-muon'' data - Monte Carlo comparison}\label{sec:dimuonsDataMC}
Given the highly penetrating power of high-energy muons, depositing only an ionization track in the ECAL, -- where they are mainly produced -- VETO and HCAL, di-muon events represent an ideal benchmark for this analysis,  featuring a signature close to that expected from the signal: a well identified 70 GeV/c track in time with a positron-like signal in the SRD and a low-energy electromagnetic shower in the ECAL, with a significant missing energy. Other than providing a signal-like data sample for the efficiency evaluation of the ECAL cuts, as described in Sec.~\ref{sec:ECALSignalMC}, di-muons are also used to estimate a global systematic uncertainty factor on the signal detection efficiency.
Specifically, the comparison between the measured di-muon energy spectrum in the ECAL and that predicted by MC simulations allows to cross-check the overall detector response, 
and to compare the absolute events normalization. 
In the setup here considered, roughly one di-muon event is expected per 10$^5$ $e^+$OT. To efficiently simulate di-muon events within Geant4~\cite{Allison:2016lfl,GEANT4:2002zbu}, we artificially biased the corresponding cross section by a factor of 200, for both the radiative process and the annihilation one. This adjustment allowed us to gather a sufficient statistics within a feasible time without distorting the simulation of the shower development in the ECAL (see e.g. Ref.~\cite{Banerjee:2017hhz} for a complete discussion regarding the implementation of the radiative process within the simulation framework).

In order to select di-muon events in data, we first applied all upstream cuts to identify events with a clean 70 GeV/c positron track  (see Sec.~\ref{sec:upstreamCuts}). Then, 
we required that the total energy deposited in the pre-shower was greater than 350 MeV and that the total deposit in the central cell of HCAL0, HCAL1, and HCAL2 was within the range  $0.5 - 9.5$ GeV, compatible with a double-MIP track. Finally, to suppress pile-up events in data, we imposed the condition E$_{ECAL}^{OOT} < 2\times$E$_{ECAL}^T$, where E$_{ECAL}^{OOT}$ (E$_{ECAL}^T$) is the out-of-time (in-time) energy deposited in the ECAL (see Sec.~\ref{sec:dataProc}). 
\begin{figure}[t]
    \centering
    \includegraphics[width=0.8\linewidth]{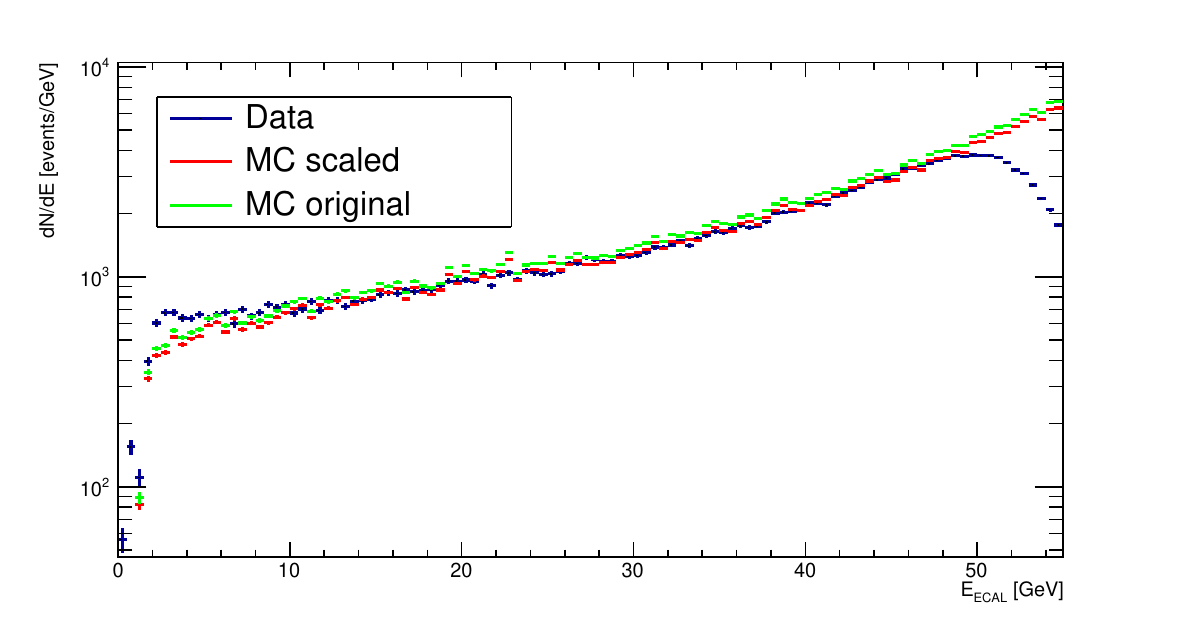}
    \caption{Comparison of the di-muons yield for data (blue) and MC. The green histogram shows the MC spectrum scaled according to Eq.~\ref{eq:MC_Scaling_Dimuons}. The red histogram represents the MC spectrum scaled for an additional factor 0.93. For data, the energy range $E_{ECAL} \gtrsim 50$~GeV shows the effect of the ECAL trigger threshold.}
    \label{fig:dimuonsDataMC}
\end{figure}
The same selection was applied to the MC, except for the upstream cuts and the pile-up cut, since neither hadron contamination nor pile-up effect were included in the simulation framework.
The comparison of the obtained ECAL spectra is reported in Fig.~\ref{fig:dimuonsDataMC}. Here, the MC distribution is scaled for a factor $f$ according to:
\begin{equation}\label{eq:MC_Scaling_Dimuons}
    f = \frac{1}{N_{e^+OT}^{MC} \times B} \times N_{e^+OT}^{data}\times \eta_{upstream}\; \times \eta_{pile-up} \; ,
\end{equation}
where $N_{e^+OT}^{MC}$ is the number of primary positrons that were simulated, $B$ is the bias factor which was used to enhance the cross section associated with the di-muons production through annihilation or nuclear interaction, $N_{e^+OT}^{data}$ is the total number of \pot acquired during the whole measurement, $\eta_{upstream}$ is the efficiency associated with the upstream cuts, and $\eta_{pile-up}$ is the efficiency of the aforementioned cut on the ECAL in-time/out-of-time energy. 

Overall, a good agreement between the two distributions is observed, both in terms of shape and normalization, confirming the robustness of the present analysis, and of proper understanding of the detector response. A residual shape discrepancy is noticeable at low energy, with the MC result under-estimating the measured data, possibly because of the effect of the ``time-cut'' correction (Eq.~\ref{eq:intime}) on residual pile-up events not fully rejected by our selection, or due to a discrepancy in the effective pre-shower threshold value between data and MC, particularly relevant for annihilation events, accumulating in the low $E_{ECAL}$ region.
The overall number of di-muon events per incident positron ($n^{\mu\mu}_{1\, e^+OT}$) reads $n^{\mu\mu}_{1\, e^+OT}$ = 6.62$\times10^{-6}$ (7.12$\times10^{-6}$) for data (MC), leading to a data-MC ratio of $\mathcal{F}^{\mu\mu} = 0.93$. This has been accounted for in the analysis as a global efficiency term for the signal, that we assumed to be possibly due to an inefficiency of the online missing-energy trigger. We assessed two sources of systematic uncertainty affecting this ratio. The first considers the long-term variations of the detector response not completely compensated by run-by-run calibrations, or the differences in the impinging beam properties (divergency and transverse size) possibly affecting the upstream detectors efficiency. This term was evaluated from the run-by-run fluctuations of $\mathcal{F}^{\mu\mu}$. To avoid double-counting effects, we subtracted the fluctuations of $\mathcal{F}^{\mu\mu}$ due to run-by-run variations in the MC ECAL energy threshold. The second term, instead, accounts for possible mismatches between data and MC for the HCAL response, affecting the di-muons selection. This was estimated by repeating the evaluation of $\mathcal{F}^{\mu\mu}$ varying for both datasets the HCAL selection ranges, and then assessing the variation of the results. The obtained uncertainty was $1.8\%$.

\subsection{Systematic uncertainties} 

A summary of the systematic uncertainty contributions affecting the $\Apr$ production yield that were considered for this analysis is illustrated in Tab.~\ref{tab:Systematic_uncertainties}. The first term is due to the uncertainty affecting the pile-up correction introduced for the total number of accumulated \pot, discussed previously in Sec.~\ref{sec:ECALSignalMC}. The second and third uncertainty terms, instead, are connected to the signal efficiency for the ECAL and for the other detectors, respectively. These were estimated by slightly varying the criteria to select the event sample used to determine it and assessing the corresponding variation. As described thoroughly in Sec.~\ref{sec:sig_eff}, for all but the ECAL cuts, the efficiency was estimated from a sample of 70 GeV/c impinging $e^+$, mainly selected through the requirement of an energy deposit in the calorimeter close to the beam energy. For ECAL cuts, instead, a clean sample of di-muon events was employed, selected mostly by cutting on the HCAL energy deposit. The uncertainty for the $\chi^2_{shower}$ correction, parameterized by the $C_{\chi^{2}\,MC}$ multiplicative factor, was evaluated from the bin-to-bin fluctuations of the latter in the low $\Apr$ mass region, where it is expected to be constant. This contribution to the signal efficiency uncertainty applies only for the $\Apr$ resonant production mechanism -- for radiative $\Apr$ emission we assumed the signal $\chi^2_{shower}$ distribution to match that of the di-muon events from which the corresponding efficiency was evaluated.
Finally, for the overall signal normalization uncertainty, we accounted for the two uncertainty terms affecting $\mathcal{F}^{\mu\mu}$, discussed in the previous section.
\begin{table}[t]
    \centering
    \begin{tabular}{l|l}
    \hline 
         Source & Value \\ \hline \hline
         \pot - pile up correction & 0.5\% \\
         Efficiency - ECAL cuts & 0.7\% \\
         Efficiency - $\chi^2_{shower}$ correction &  0.6\%\footnote{This contribution applies only to the $e^+e^-$ resonant annihilation channel for $\Apr$ production. }\\
         di-muon data-MC comparison & 1.7\% $\oplus$ 1.8\% \\
         \multirow{2}{*}{ECAL threshold} & Mass dependent: \\
         & max  $\simeq 4\%$ for $m_\Apr=170$~MeV. \\
         \multirow{2}{*}{$\Apr$-strahlung yield} & Mass dependent: \\
         &  6.5\% $<$ $\Delta$ $<$ 7.6\% for 1 MeV $<$ $m_{\Apr}$ $<$ 100 MeV \\
         
         \hline
    \end{tabular}
    \caption{\label{tab:Systematic_uncertainties}  Summary of the systematic uncertainties affecting the $\Apr$ yield as discussed in the text. 
    The contribution coming from the uncertainty in the efficiency of the non-ECAL cuts (described in Section~\ref{sec:UpstreamDownstreamEff}) is not included in this table because it is considered negligible. }
\end{table}

All the previous effects are independent from the $\Apr$ mass value. In contrast, the systematic uncertainty arising from the ECAL energy scale results in a mismatch between data and MC, affecting the pre-shower and ECAL thresholds, and determines a $m_\Apr$-dependent effect. We estimated the threshold uncertainties repeating the corresponding evaluation independently for each data-taking run, and evaluating the corresponding fluctuation. The obtained values where, respectively, $7\%$ for the pre-shower and $0.4\%$ for the ECAL. We thus repeated, for each value of $m_\Apr$, the calculation of the event yield within the signal window by varying the corresponding cuts within these ranges. This resulted in a worst-case relative yield variation of 4\% (3\%) for $\alpha_D = 0.1$ (0.5), for a dark photon mass of 170 MeV. We observe that this value corresponds to a resonant energy of about 28 GeV, close to the missing energy threshold. Finally, we took into account a possible systematic uncertainty associated with the computed value of the $\Apr$-strahlung cross section implemented in the DMG4 package, as described in~\cite{Gninenko:2017yus}. This contribution was found to be modestly dependent on the dark photon mass, ranging from 6.5$\%$ for $m_\Apr=1$~MeV to $7.6\%$ for $m_\Apr=100$~MeV -- a linear parameterization with respect to $log(m_\Apr)$ was adopted to cope with this.

\subsection{Results}\label{sec:res70GeV}

\begin{figure}[t]
    \centering
    \includegraphics[width=0.7\linewidth]{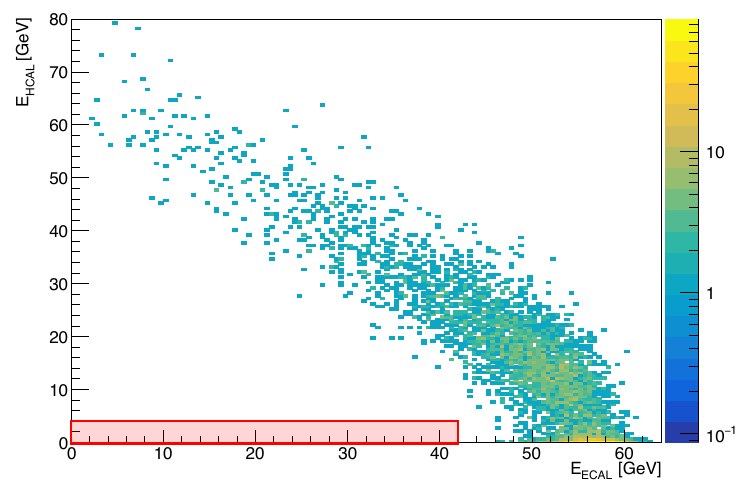}
    \caption{Events distribution in the $E_{HCAL}$ vs $E_{ECAL}$ space after the all selection cuts. The red rectangle highlights the signal region (the Y-axis extent is enlarged by a factor of 4 for better visibility).}
    \label{fig:HermeticityPlot}
\end{figure}

The un-blinded $E_{HCAL}$ vs $E_{ECAL}$ distribution of events surviving all but the HCAL energy cut is shown in Fig.~\ref{fig:HermeticityPlot}\footnote{The experiment trigger system computes the ECAL energy summing the raw amplitudes from the detector and compares the results with a fixed hardware threshold. As a result, the trigger threshold curve measured as a function of the offline-reconstructed ECAL energy shows a finite width of about 1.1 GeV as shown in Fig.~\ref{fig:enh} (top-right). When convoluted with the “beam-only” ECAL energy distribution, that shows a sharp rise toward 70 GeV, the resulting spectrum shows a tail extending above the 52 GeV nominal trigger threshold value.}. In the residual events the impinging 70 GeV/c positron produced secondaries in the ECAL, that leaked from the latter and were subsequently intercepted by the HCAL. No events were observed in the signal region, highlighted by the red band in the Fig.. 

Based on this null result, we derived the upper limit on the $\Apr$ kinetic mixing parameter $\varepsilon$ as a function of $m_\Apr$. We adopted a frequentist approach, considering the $90\%$ Confidence Level (CL) of a one-sided profile-likelihood test statistics~\cite{Gross:2007zz}. The likelihood model was built assuming a Poisson PDF for the number of events in the signal region, with mean value $\mu=S+B=(\varepsilon/\varepsilon_0)^2S_0+B$. Here, $S_0$ is the signal yield for the nominal coupling value $\varepsilon_0$ obtained from MC, taking also into account the signal efficiency, while $B$ is the expected number of background events within the signal region.  
The systematic uncertainties were accounted for in the procedure by introducing, for each contribution, an additional log-normal PDF term in the Likelihood, taking the measured value as the \textit{observed} one, and handling the \textit{expected} value as a nuisance parameter~\cite{Gross:2007zz}. These include the $\Apr$-yield uncertainty factors discussed in the previous section, as well as the $\simeq 40\%$ uncertainty for the expected number of background events. The limit calculation was performed using the COMBINE tool developed by the CMS collaboration~\cite{CMS:2024onh}, for discrete values of $m_\Apr$ in the range 1 MeV - 1 GeV.

\begin{figure}[t]
    \centering
    \includegraphics[width=0.48\linewidth]{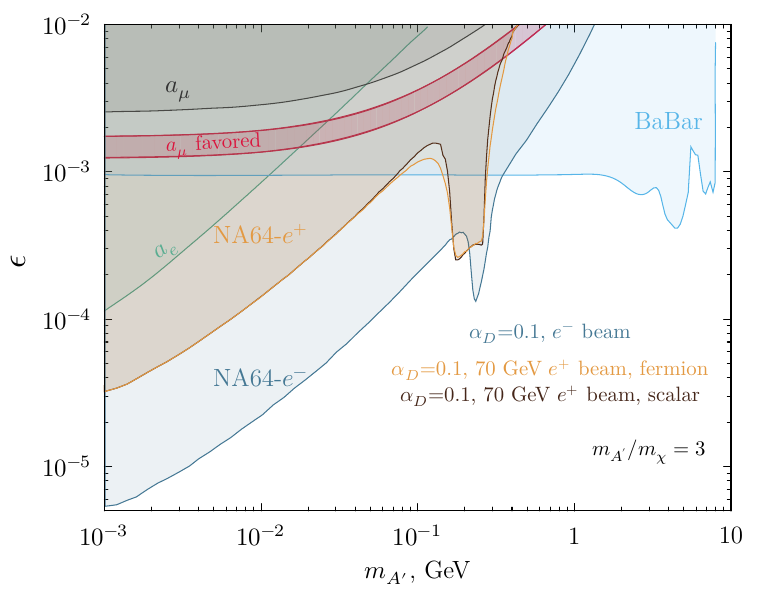}
    \includegraphics[width=0.48\linewidth]{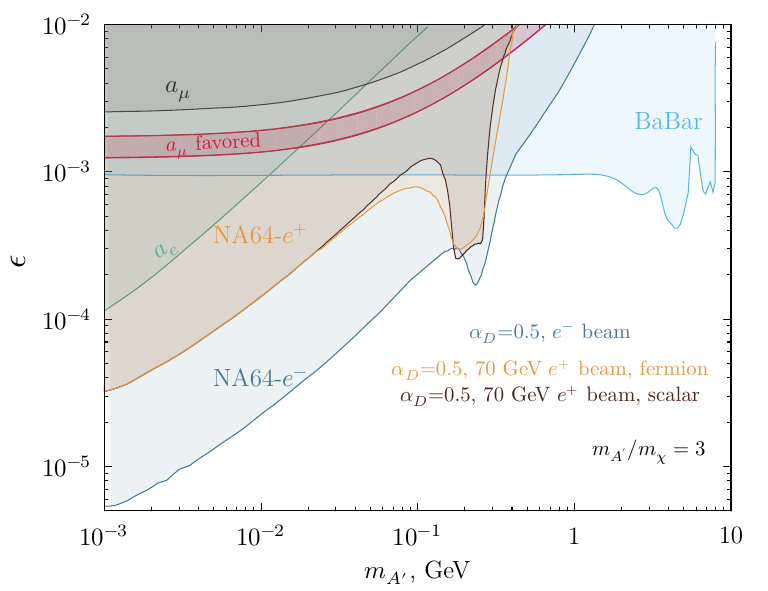}
    \includegraphics[width=0.48\linewidth]{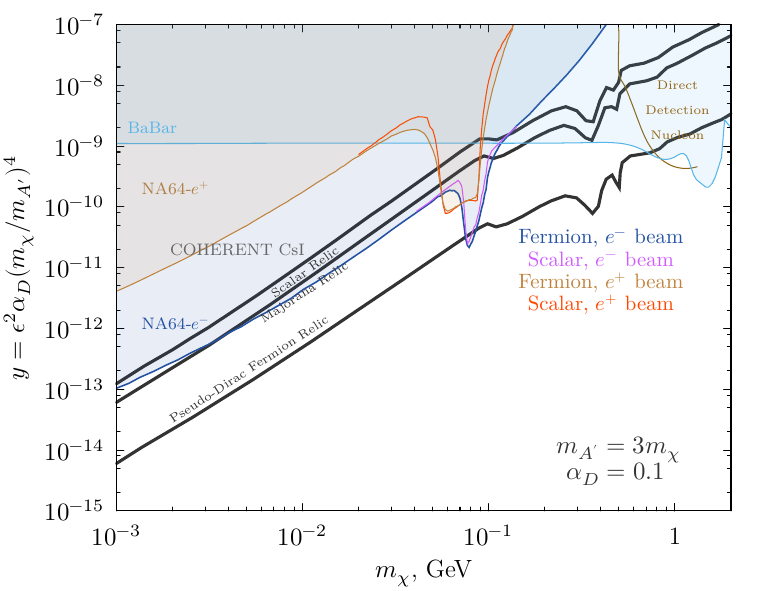}
    \includegraphics[width=0.48\linewidth]{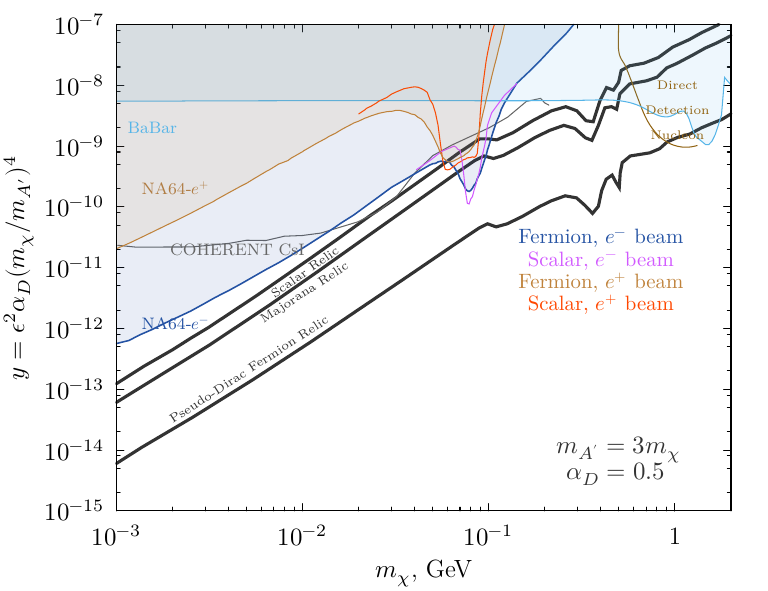}
    \caption{Exclusion limits at 90\% confidence level derived from the 70 GeV/c positron-beam missing energy measurement presented in this work. Top (bottom): exclusion limits - brown and orange curves- in the 
    $[m_\Apr,\varepsilon]$ ($[m_\chi,\alpha_D]$) plane, for $\alpha_D$ = 0.1 (left) and $\alpha_D$ = 0.5 (right). The other curves and shaded areas report already-existing limits in the same parameters space from NA64 in electron-beam mode~\cite{Andreev:2023uwc} (blue and violet), COHERENT~\cite{COHERENT:2021pvd} (grey), and BaBar~\cite{BaBar:2016sci} (light blue). In the bottom plots, the black lines show the favored parameter combinations for the observed dark matter relic density for different variations of the model.}
    \label{fig:results70GeV}
\end{figure}

The results are depicted in Fig.~\ref{fig:results70GeV}, in terms of exclusion curves in the $[m_\Apr,\varepsilon]$ space (top row) and in the $[m_\chi, y]$ one. In the latter, the black lines represent various cosmological targets, each corresponding to a different thermal origin of DM. To check the dependency of our result with respect to the LDM model parameters, we repeated the upper limit calculation for the two DM coupling constant values $\alpha_D=0.1$ and $\alpha_D=0.5$, and considering both a scalar and a fermion model for DM.

In the exclusion limit curves, the contribution of the resonant production mechanism is clearly indicated by the enhancements located between 100 and 300 MeV. The shape of the resonant peak consists of two ``lobes''. The right lobe is a result of the convolution of the $e^+$ track length in the ECAL with the resonant peak. The left lobe is due to the signal box upper limit in the ECAL ($0 < E_{ECAL} <$ 42 GeV). Despite the limited acquired statistics, thanks to the strong enhancement of the signal yield due to the $e^+e^-$ annihilation mechanism, particularly effective for a primary positron beam, we were able to explore a new region in the LDM parameter space, while maintaining control over all background sources. 
Specifically, our results exclude the existence of vector-mediated light dark matter in the mass range $165 < m_\Apr < 220$ MeV, for $\varepsilon$ values down to $2.5\times10^{-4}$ and $\alpha_D=0.1$, and corroborates the already-observed exclusion contours for $\alpha_D=0.5$ in the same window by the NA64-$e^-$ run, with a factor $\sim100\times$ larger accumulated statistics.

\section{Future projections}

\begin{figure}
    \centering
    \includegraphics[width=0.8\linewidth]{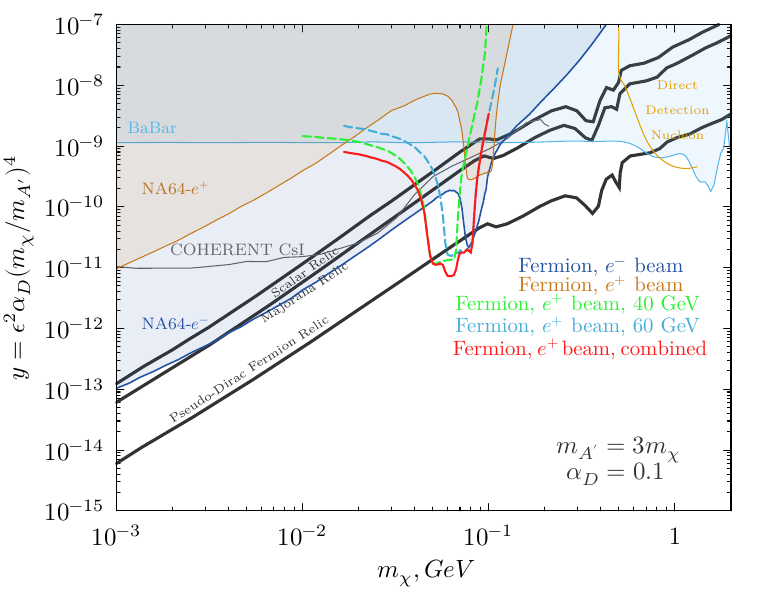}
    \caption{Projected sensitivity of the proposed positron measurements in the $(y,m_\chi)$ plane in the $(m_\chi,y)$ plane, for $\alpha_D=0.1$. The green and cyan curves are the results obtained from the detailed simulation of the 40 GeV and 60 GeV measurement, and the red curve corresponds to their combination. }
    \label{fig:addendum}
\end{figure}

 The strong enhancement to the $\Apr$ signal yield associated with the $e^+e^-$ resonant annihilation production channel allows a limited-statistics positron-beam missing energy experiment to efficiently probe the LDM parameter space down to the relic density targets for the dark photon mass range defined by Eq.~\ref{eq:A_mass_range}. The preliminary sensitivity for
 a dedicated program at NA64 with two runs at different beam energies, 60 GeV and 40 GeV, was first estimated in Ref.~\cite{NA64:2023ehh}, assuming zero expected background event in the signal window, defined by the condition $E_{ECAL}<E_{beam}/2$. The result is highlighted in Fig.~\ref{fig:addendum}, showing the expected sensitivity for a future program foreseeing two data-taking runs at the aforementioned energies, with $10^{11}$ \pot accumulated for each configuration. This effort will allow to probe the LDM parameter space down to the Pseudo-Dirac fermion relic curve, with a reduced statistics corresponding to just few weeks of run.

The measurement presented in this work, based on a 70 GeV/c $e^+$ beam configuration and a missing energy threshold $E_{th}=28$ GeV, allows us to experimentally validate the sensitivity of the proposed run at 60 GeV. Indeed, thanks to the reduced $E_{th}$ value, the present configuration covers the $\Apr$ mass range $170~\mathrm{MeV}\lesssim m_\Apr \lesssim 270$~MeV, similar to what expected for a 60 GeV/c measurement with $50\%$ missing-energy threshold. Considering the observed background yield of $\simeq 5.4\times 10^{-12}$ events/\pot, the conservative estimate for the full statistics of $10^{11}$ \pot reads $B\simeq 0.5$ events, strongly dominated by events in which the primary $e^+$ interacts with upstream beam line materials and gives rise to secondary hadrons falling out of the detector acceptance. This estimate is supported by the observation that no significant variations for the total electro-/photo-nuclear cross section are present between the two regimes, and the kinematics of final state products is similar~\cite{ParticleDataGroup:2024cfk}.

Concerning the 40 GeV/c measurement, instead, a first evaluation of the expected backgrounds was discussed in Ref.~\cite{Bisio:2887649}. While the contribution from hadrons and muons in-flight decay is further suppressed by the reduced beam contamination, of the order of $\eta_h \simeq 0.03\%$, the background yield due to upstream interactions was found to be a factor $\simeq \times 10$ larger, mostly because of the larger emission angle of secondary particles, resulting in a lower detection acceptance. This calls for an optimization of the experimental setup. 

As a first investigation, we considered an improved geometrical alignment of the existing NA64$e$ detectors, with the VHCAL prototype moved close to the ECAL. 
We executed a FLUKA-based simulation of this experimental setup, implementing an artificial bias factor for the cross section of electro- and photo-nuclear reactions\footnote{Biasing was activated only for the detector volumes located upstream the ECAL.} in order to enhance the statistics. The final result read $\simeq 0.1$ background events for the accumulated statistics of 10$^{11}$ \pot-- for further details regarding the validation of this simulation, using the 70 GeV/c dataset, see Appendix~\ref{app:upstremInteractionsSimuDataComparison}.
This confirms the feasibility of the low-energy measurement and the validity of the zero background assumption adopted in the sensitivity evaluation for an accumulated statistics of $10^{11}$\pot. 
The result also demonstrates that, in view of a further extension of the NA64e program, probing the thermal relic targets for $\alpha_D > 0.1$, a further change of the setup is required, to keep the background source under control. To this end, the NA64 collaboration is currently working toward the construction and the installation of a full-scale VHCAL detector, with optimazed angular coverage~\cite{vhcalBenjamin}.

\section{Conclusions}
In this manuscript we present the analysis of the $1.596\times 10^{10}$ \pot measurement collected by NA64$e$ in a $\sim$24-hours-long measurement during summer 2023, with a 70 GeV/c positron beam. This preliminary effort, representing the first physics measurement of NA64$e$ with a beam energy lower than 100 GeV, was conducted as a first step towards a possible future program, involving positron beams with energies down to 40 GeV. We adopted a blind approach for the analysis, optimizing the event selection criteria based on their impact on the signal detection efficiency vs background rejection power, with particular attention to the optimization of the detection efficiency for signals induced by positron resonant annihilation. Different strategies were adopted depending on the cut nature, favoring, where possible, a data-driven approach. Similarly, the main expected backgrounds of the experiment were evaluated with a mix of data-driven and MC-based techniques, resulting in a overall estimate of $(0.08\pm 0.07_{stat} \pm 0.03_{sys})$ background events. No events were found in the un-blinded signal window, resulting in new constraints in the LDM parameter space: despite the limited accumulated statistic, a small unexplored region was explored for specific variations of the model, proving the robustness of the technique. 

In addition to the physics result, this analysis provided crucial information towards the proposed positron program at NA64: the low background projections strongly support the feasibility of a full zero-background 60 GeV/c measurement, collecting up to $10^{11}$ \pot; on the other side, the data-MC comparison of the upstream nuclear interactions is critical for a future 40-GeV/c program, since these processes can contribute significantly to the expected background for such measurement; the obtained results indicated a mild underestimation of the rate of these events in MC simulations, motivating a conservative approach in the background extrapolations. 

 \appendix

\section{Study of the di-muon - resonant annihilation  shower shape comparison}\label{app:chi2_correction}
The equivalence, from the point of view of the ECAL energy deposit, of di-muon and signal events, is a crucial aspect of the analysis, in particular for the study of the efficiency of the $\chi^2_{shower}$ cut described in Sec.~\ref{sec:downstream_cuts}. This cut, optimized to get an efficiency of $95\%$ for di-muon events, is the most sensitive, among the ECAL-based selections, to the development of the electromagnetic shower within the ECAL, requiring an in depth-analysis of its effect on signal events. The study here described, in particular, aims to investigate the equivalence, from the point of view of the $\chi^2_{shower}$ cut, of di-muon events and signal events where a  LDM particles pair is produced via resonant annihilation, being this the main LDM production process of interest in this work. In order to avoid any bias in the direct comparison of data and MC simulations, possibly induced by imperfections in the MC description of the ECAL response, the study was performed only by means of simulations of both di-muon and $\Apr$ resonant annihilation production events. As a first step, a reference ``MC profile'' was built using simulated 70 GeV/c positron events. The profile was obtained with a procedure analogous to that adopted for profiles produced with real data. Then, a set of di-muon events was produced via simulations (see Sec.~\ref{sec:dimuonsDataMC} for details on the di-muons MC simulation), and the $\chi^2_{shower}$ distribution was computed for this set, using the aforementioned MC profile as a reference.  Following the procedure adopted for data, the $\chi^2_{shower}$ distribution obtained for di-muons was used to optimize the  $\chi^2$ cut values, in different energy ranges; the values were selected to have a flat $95\%$ efficiency over the 0-55 GeV ECAL energy range. 
Finally, once the cut was defined, it was applied to different simulated sets of signal events, where the LDM particle pair was produced via resonant annihilation. The different signal sets were produced by varying the mass of the $\Apr$ in the 1 MeV - 350 MeV interval, to cover the range of interest for this measurement; four different simulation sets, varying $\alpha_D$ between 0.1 and 0.5 and testing both scalar and fermion LDM particles, were produced. Only events where the LDM pair carried away significant energy (less than $\sim60$ GeV deposited in the ECAL) were saved, since events with lower missing energy would not be written to disk given the trigger definition. The efficiency of the shower cut was then evaluated for each of the configuration/$m_\Apr$ values simulated. To account for the pre-shower trigger effect, the study was performed by selecting only events surviving a 350 MeV cut on the pre-shower energy. The additional requirement of $1 \,\,{\rm GeV}<E_{ECAL}<42\,\,{\rm GeV}$ was applied, in order to focus on  LDM events falling in the signal box, with a large enough ECAL energy deposit for the calculation of $\chi^2_{shower}$. Indeed, for events with $E_{ECAL}<1\,\,{\rm GeV}$, the shower cut efficiency is fixed to zero, since the energy is deposited almost completely in one cell, making meaningless the $\chi^2_{shower}$ variable\footnote{In the analysis, this affect was accounted for in the evaluation of the overall signal efficiency of the ECAL cuts.}. 
\begin{figure}
    \centering
    \includegraphics[width=.9\textwidth]{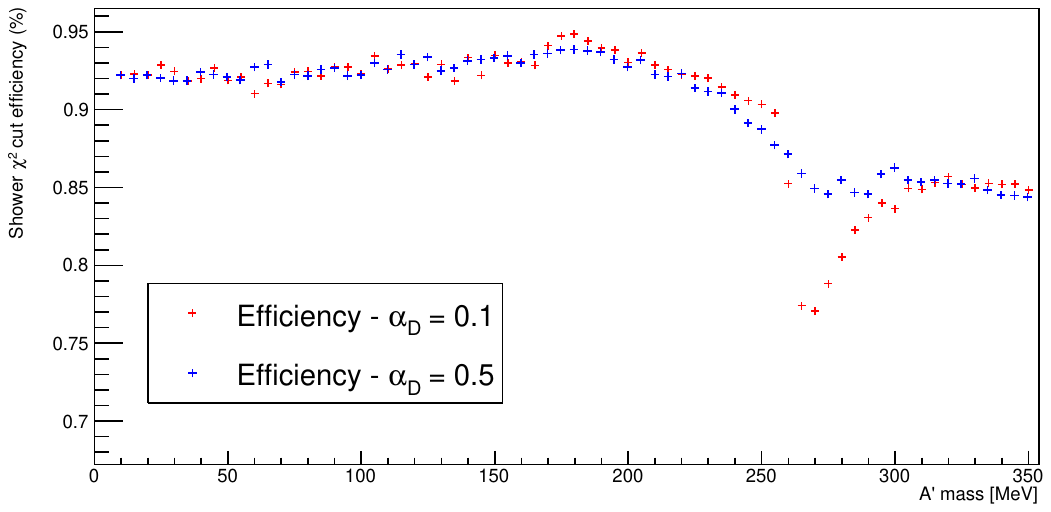}
    \caption{Red (blue) markers: efficiency of the shower shape cut on resonant annihilation events simulated with $\alpha_D =0.1$ ($\alpha_D =0.5$), as a function of the $\Apr$ mass.}
    \label{fig:eff_comp}
\end{figure}
Fig.~\ref{fig:eff_comp} shows the result of this survey: the red (blue) markers show the efficiency of the shower shape cut on resonant  annihilation events simulated considering a value of $\alpha_D =0.1$ ($\alpha_D =0.5$) for a fermionic LDM particle candidate, as a function of the $\Apr$ mass. The efficiency is approximately flat, between $92\%$ and $94\%$, up to the $\sim$ 200 MeV mark, where it drops, reaching a minimum at $m_\Apr \sim 270$ MeV. This effect is due to resonant energy $E_{res} = \frac{m_\Apr^2}{2m_e}$ approaching the primary energy of the beam as $m_\Apr \rightarrow$ 270 MeV. Indeed, an $\Apr$ with such mass can be produced resonantly from a positron of about 70 GeV, corresponding to the primary beam energy. For $m_\Apr$ values approaching $\sim$270 MeV, therefore, the  $\Apr$ is mainly produced by the primary positron in the first centimeters of the ECAL, resulting in negligible energy deposited in the ECAL, and in a low efficiency of the shower shape cut. An analogous effect can be observed for the simulations assuming a scalar LDM candidate.

This study suggests that the optimization procedure of the $\chi^2_{shower}$ cut  with di-muons results in a slight loss of efficiency on resonant annihilation events, for masses larger than $\sim 200$ MeV. In other words, di-muon events can't be considered equivalent to resonant annihilation events independently of the mass of the $\Apr$. In conclusion, we decided to take into account this effect in the positron analysis by correcting the signal efficiency with a factor depending on the mass $m_{\Apr}$, obtained by extrapolation from the graphs shown in Fig.~\ref{fig:eff_comp}. A corresponding systematic error of $Err_{\chi^2}\simeq 0.6 \%$ was estimated by considering the fluctuations of the values shown in the graphs with respect to a moving average over 5 points.

\section{Improved $\Apr$ production simulations in Geant4 via \texttt{G4StackingAction}}\label{app:stackingAction}

To optimize computation time, the Geant4 NA64 simulation framework for $\Apr$ events simulation, based on the DMG4 package, includes a ad-hoc biasing for the production cross section. This artificial enhancement allows to increment the number of events in which an $\Apr$ is generated by the interaction of the primary positron or any of the secondaries within the electromagnetic shower developing in the ECAL with the materials of the latter. Still, as discussed in Sec.~\ref{sec:ECALSignalMC}, even with this enhancement the fraction of events featuring dark-photon production cannot exceed $\mathcal{O}(\%)$ of the total, to avoid non-linearities in the simulation. As a result, the large majority of the computation time is spent for events that are of no interest in the calculation of the signal yield and corresponding efficiency. Evidently, introducing for all events a cut on the energy or range of secondary particles in the shower is not a solution for this, since it would bias the ECAL response also for signal-like events.

To overcome this limitation, we introduced an optimized simulation scheme. This is based on the observation that, given the NA64 missing-energy condition, in signal-like events the $\Apr$ must be produced by the primary $e^+$ or any of the secondaries within the first few ECAL radiation lengths, at the beginning of the shower development, to have enough energy to be detected. Only for these events a full simulation of the shower development is actually required. For this, we exploited the existing Geant4 \texttt{G4UserStackingAction} class, that allows to selectively simulate low-energy particles depending on the effective $\Apr$ production. This is made possible by introducing two particles stacks, \texttt{fWaiting} and \texttt{fUrgent}. Any \texttt{G4Track}  in the \texttt{fUrgent} stack is simulated immediately, in the order it is pushed to the stack, while those in the \texttt{fWaiting} stack are not simulated until the \texttt{fUrgent} one is empty.

Specifically, the \texttt{G4UserStackingAction} class introduces three methods, to be implemented in an application-specific class inheriting from it.
\begin{itemize}
\item The \texttt{ClassifyNewTrack()} method is called every time a new track is to be simulated, either the primary one (at the beginning of each event) or a secondary one. This method returns an enumerator, \texttt{G4ClassificationOfNewTrack()}, that indicates to which stack, if any, the track will be sent. This method is invoked for each \texttt{G4Track} created in the event, including the primary one.
\item The \texttt{NewStage()} method is invoked when the \texttt{fUrgent} stack is empty and the \texttt{fWaiting} one contains at least one \texttt{G4Track} object. In the default version, it will move all objects in the \texttt{fWaiting} stack (if any) to the \texttt{fUrgent} one.
\item The \texttt{PrepareNewEvent()} method is invoked at the beginning of each event, when all stacks are empty.
\end{itemize}

Our implementation works by first simulating, for each event, all high-energy particles in the ECAL, that could possibly produce an $\Apr$ particle, while ``pausing'' the simulation of the low-energy secondaries, through appropriate use of the two stacks mentioned before. In case the $\Apr$ production actually occurs, the latter are also simulated.
\begin{itemize}
\item The \texttt{ClassifyNewTrack()} method determines to which stack any new \texttt{G4Track} is to be loaded based on the value of a \texttt{stage} flag. When this is zero, all \texttt{G4Tracks} with energy greater than a programmable threshold $E_{min}$ are loaded on the \texttt{fUrgent} stack, and the others to the \texttt{fWaiting} one. When the flag is one, all \texttt{G4Tracks} are loaded to the \texttt{fUrgent} stack.
\item The \texttt{PrepareNewEvent()} method initializes the \texttt{stage} flag to zero. 
\item In the \texttt{NewStage()} method, the \texttt{stage} flag is set to one if any $\Apr$ production occurred in the event, actually forcing the complete simulation of the latter. Otherwise, the event simulation is terminated via a call to \texttt{stackManager->clear()}.
\end{itemize}

\section{Validation of the FLUKA-based simulation of the upstream electro/photo nuclear interactions at 40 GeV}\label{app:upstremInteractionsSimuDataComparison}
\begin{figure}
    \centering
    \includegraphics[width=0.8\linewidth]{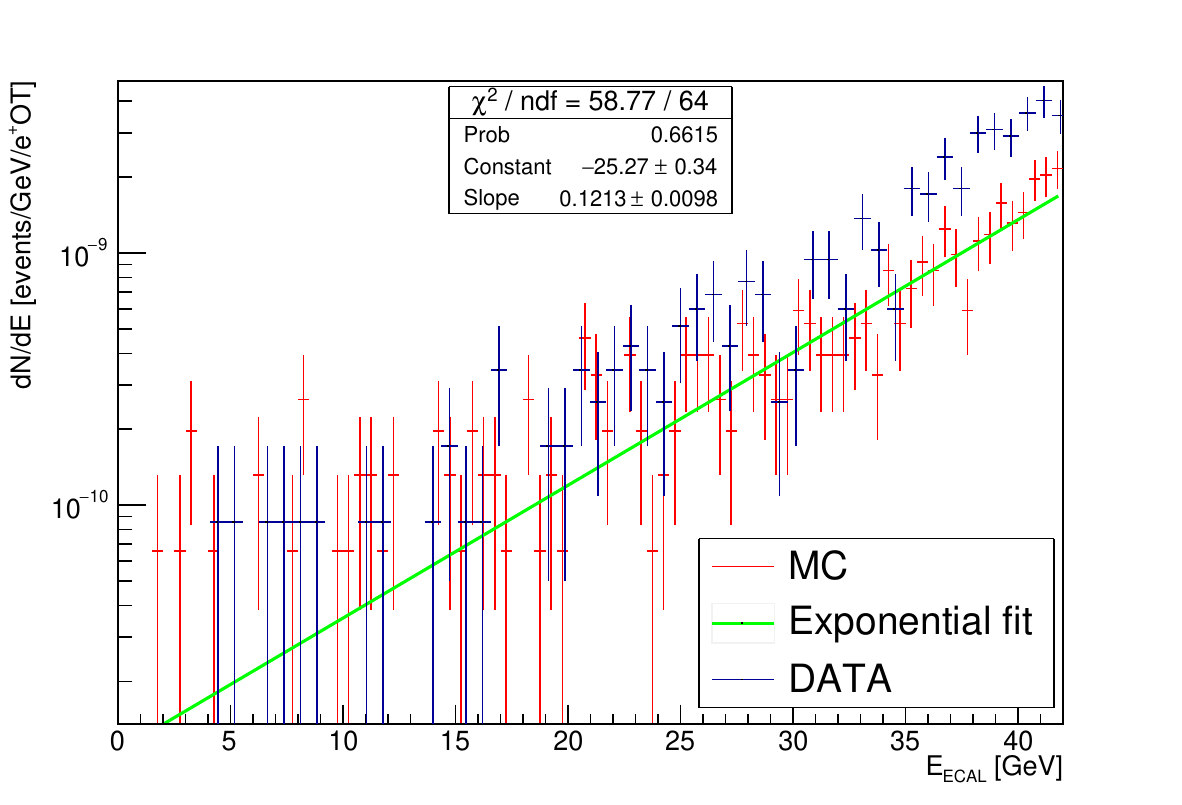}
    \caption{ECAL energy distribution for selected electro-/photo-nuclear upstream events: data (blue) vs MC prediction (red). The MC distribution is fitted with an exponential function $f(E_{ECAL})=\exp(-c_{const}+c_{slope}E_{ECAL})$ (in green) to assess the systematic uncertainty connected to the extrapolation of the upstream-interactions background yield(see Sec.~\ref{sec:upstreamInteractions} for details). }
    \label{fig:MCcomparison}
\end{figure}

In this section, we delved into the study of upstream electro/photo-nuclear interactions, which serve as the primary background sources for this analysis. This study holds significant importance, particularly in light of the potential future NA64$e$ program that involves positron beams with energies as low as 40 GeV.
In this context, we conducted MC simulations using the FLUKA toolkit and utilized the current 70 GeV/c measurement to validate the accuracy of these simulations.
We concentrated on events involving the primary electron positron ($e^+$) and upstream beam line materials, specifically those that result in elector/photo-nuclear interactions. In the dataset, we isolated these by applying all the selection cuts discussed in Sec.~\ref{sec:cuts}, apart from the ECAL $\chi^2_{shower}$. Also, to enhance statistics, we decided to exclude the VHCAL cut from the analysis. The experimental ECAL energy distribution was compared to the MC prediction, both normalized to a single \pot. We observe a good agreement in terms of distribution shape, and also for the overall normalization (see Fig.~\ref{fig:MCcomparison}). The total number of events in the data sample reads $(29\pm1)\times 10^{-9}$ events/$e^+$OT, while the MC predicts $(15.7\pm0.7)\times 10^{-9}$ events/\pot. This $\simeq \times 2$ discrepancy in the overall number of events is possibly due to a mismatch in the parameterization of the cross sections for these rare processes in MC, or to an partially inaccurate modeling of the beam properties and on the composition of all upstream beam line materials. 
To assess the dependency of the MC results on the parameterization of the beam properties (angular divergence and transverse size), we repeated the calculation considering an ideal ``pencil-like'' configuration, obtaining similar results.
In conclusion, even accounting for the factor $\times2$ discrepancy, the FLUKA background estimate for the 40 GeV/c measurement, performed with the same simulation framework, predicts $\simeq 0.1$ events for the total statistics of $10^{11}$ \pot.

\acknowledgments

We gratefully acknowledge the support of the CERN management and staff, and the technical staffs of the participating institutions for their vital contributions. 
This result is part of a project that has received funding from the European Research Council (ERC) under the European Union's Horizon 2020 research and innovation programme, Grant agreement No. 947715 (POKER). 
This work was supported by the HISKP, University of Bonn (Germany), ETH Zurich Grant No. 22-2 ETH-031, and SNSF Grant No. 186181, No. 186158, No. 197346, No. 216602 (Switzerland), and FONDECYT (Chile) under Grant No. 1240066 and Grant No. 3230806, and ANID - Millenium Science Initiative Program - ICN2019 044 (Chile), and  RyC-030551-I and PID2021-123955NA-100 funded by MCIN/AEI/ 10.13039/501100011033/FEDER, UE (Spain), and COST Action COSMIC WISPers CA21106, supported by COST (European Cooperation in Science
and Technology). This work is partially supported by ICSC – Centro Nazionale di Ricerca in High Performance Computing, Big Data and Quantum Computing, funded by European Union – NextGenerationEU.



\bibliographystyle{JHEP}
\bibliography{bibliographyNA64_inspiresFormat,bibliographyNA64exp_inspiresFormat,bibliographyOther_inspiresFormat}

\end{document}